\let\TeXyear\year
\documentclass[usenames,dvipsnames]{ieeeaccess}

\let\year\TeXyear
\usepackage{amsmath,amssymb,amsmath,amsfonts}
\usepackage{bm}
\usepackage{graphicx}
\usepackage{cite}
\usepackage{epstopdf}
\usepackage{balance}
\usepackage{caption}
\usepackage{gensymb}
\usepackage{color}
\usepackage{lipsum}
\usepackage{float}
\usepackage{epstopdf}
\usepackage[mathcal]{eucal}
\usepackage{subfiles}
\usepackage{subcaption}
\usepackage[percent]{overpic}
\usepackage[T1]{fontenc}
\usepackage{url}
\usepackage{svg}
\usepackage{xspace}
\usepackage{siunitx}
\usepackage{ragged2e}
\usepackage{colortbl}
\usepackage{textcomp}
\usepackage{algorithmic}
\usepackage{tikz}
\usetikzlibrary{chains,arrows,calc,positioning}
\usetikzlibrary{shapes.multipart,decorations.pathreplacing}

\usepackage{tabulary}

\def\BibTeX{{\rm B\kern-.05em{\sc i\kern-.025em b}\kern-.08em
    T\kern-.1667em\lower.7ex\hbox{E}\kern-.125emX}}

\newcommand{\tblhdr}[1]{\cellcolor{BlueGreen!50} \textbf{#1}}
\newcommand{\tblsubhdr}[1]{\cellcolor{SkyBlue!15} \textbf{#1}}

\usepackage{soul}

\usepackage{breqn}
\usepackage{booktabs}
\usepackage{multirow}
\usepackage{makecell}
\newcommand{\VISION}{VIS$^4$ION\xspace}

\usepackage{soulutf8}

\usepackage[acronyms,nonumberlist,nopostdot,nomain,nogroupskip]{glossaries}
\newacronym{quic}{QUIC}{Quick UDP Internet Connections}
\newacronym{3gpp}{3GPP}{3rd Generation Partnership Project}
\newacronym{adc}{ADC}{Analog to Digital Converter}
\newacronym{5g}{5G}{fifth generation}
\newacronym{aimd}{AIMD}{Additive Increase Multiplicative Decrease}
\newacronym{am}{AM}{Acknowledged Mode}
\newacronym{amc}{AMC}{Adaptive Modulation and Coding}
\newacronym{aqm}{AQM}{Active Queue Management}
\newacronym{awgn}{AGWN}{Additive White Gaussian Noise}
\newacronym{afd}{AFD}{Austin Fire Department}
\newacronym{balia}{BALIA}{Balanced Link Adaptation}
\newacronym{bdp}{BDP}{Bandwidth-Delay Product}
\newacronym{bf}{BF}{Beamforming}
\newacronym{cc}{CC}{Congestion Control}
\newacronym{cdf}{CDF}{Cumulative Distribution Function}
\newacronym{cn}{CN}{Core Network}
\newacronym{cqi}{CQI}{Channel Quality Information}
\newacronym{cp}{CP}{Control Plane}
\newacronym{csirs}{CSI-RS}{Channel State Information - Reference Signal}
\newacronym{dc}{DC}{Dual Connectivity}
\newacronym{dce}{DCE}{Direct Code Execution}
\newacronym{dci}{DCI}{Downlink Control Information}
\newacronym{dl}{DL}{Downlink}
\newacronym{dmr}{DMR}{Deadline Miss Ratio}
\newacronym{dmrs}{DMRS}{DeModulation Reference Signal}
\newacronym{e2e}{E2E}{End-to-End}
\newacronym{ecn}{ECN}{Explicit Congestion Notification}
\newacronym{edf}{EDF}{Earliest Deadline First}
\newacronym{enb}{eNB}{evolved Node Base}
\newacronym{epc}{EPC}{Evolved Packet Core}
\newacronym{es}{ES}{Edge Server}
\newacronym{fdma}{FDMA}{Frequency Division Multiple Access}
\newacronym{fdd}{FDD}{Frequency Division Duplexing}
\newacronym[firstplural=Radio Access Technologies (RATs)]{rat}{RAT}{Radio Access Technology}
\newacronym{fs}{FS}{Fast Switching}
\newacronym{ftp}{FTP}{File Transfer Protocol}
\newacronym{gnb}{gNB}{Next Generation Node Base}
\newacronym{harq}{HARQ}{Hybrid Automatic Repeat reQuest}
\newacronym{hetnet}{HetNet}{Heterogeneous Network}
\newacronym{hh}{HH}{Hard Handover}
\newacronym{hol}{HOL}{Head-of-Line}
\newacronym{ia}{IA}{Initial Access}
\newacronym{imt}{IMT}{International Mobile Telecommunication}
\newacronym{iot}{IoT}{Internet of Things}
\newacronym{los}{LOS}{Line of Sight}
\newacronym{lte}{LTE}{Long Term Evolution}
\newacronym{m2m}{M2M}{Machine to Machine}
\newacronym{mac}{MAC}{Medium Access Control}
\newacronym{mc}{MC}{Multi-Connectivity}
\newacronym{mcs}{MCS}{Modulation and Coding Scheme}
\newacronym{mec}{MEC}{Mobile Edge Cloud}
\newacronym{mi}{MI}{Mutual Information}
\newacronym{mimo}{MIMO}{Multiple Input, Multiple Output}
\newacronym{mmwave}{mmWave}{millimeter wave}
\newacronym{mr}{MR}{Maximum Rate}
\newacronym{mss}{MSS}{Maximum Segment Size}
\newacronym{mtd}{MTD}{Machine-Type Device}
\newacronym{mtu}{MTU}{Maximum Transmission Unit}
\newacronym{nfv}{NFV}{Network Function Virtualization}
\newacronym{nlos}{NLOS}{Non Line of Sight}
\newacronym{nr}{NR}{New Radio}
\newacronym{ofdm}{OFDM}{Orthogonal Frequency Division Multiplexing}
\newacronym{pdcch}{PDCCH}{Physical Downlonk Control Channel}
\newacronym{pdcp}{PDCP}{Packet Data Convergence Protocol}
\newacronym{pdsch}{PDSCH}{Physical Downlink Shared Channel}
\newacronym{pdu}{PDU}{Packet Data Unit}
\newacronym{pf}{PF}{Proportional Fair}
\newacronym{pgw}{PGW}{Packet Gateway}
\newacronym{phy}{PHY}{Physical}
\newacronym{pbch}{PBCH}{Physical Broadcast Channel}
\newacronym[plural=\gls{mme}s,firstplural=Mobility Management Entities (MMEs)]{mme}{MME}{Mobility Management Entity}
\newacronym{prb}{PRB}{Physical Resource Block}
\newacronym{pss}{PSS}{Primary Synchronization Signal}
\newacronym{pucch}{PUCCH}{Physical Uplink Control Channel}
\newacronym{pusch}{PUSCH}{Physical Uplink Shared Channel}
\newacronym{rach}{RACH}{Random Access Channel}
\newacronym{ran}{RAN}{Radio Access Network}
\newacronym{red}{RED}{Robotics Emergency Deployment}
\newacronym{rf}{RF}{Radio Frequency}
\newacronym{rlc}{RLC}{Radio Link Control}
\newacronym{rlf}{RLF}{Radio Link Failure}
\newacronym{rrc}{RRC}{Radio Resource Control}
\newacronym{rrm}{RRM}{Radio Resource Management}
\newacronym{rr}{RR}{Round Robin}
\newacronym{rs}{RS}{Remote Server}
\newacronym{rsrp}{RSRP}{Reference Signal Received Power}
\newacronym{rss}{RSS}{Received Signal Strength}
\newacronym{rtt}{RTT}{Round Trip Time}
\newacronym{rw}{RW}{Receive Window}
\newacronym{rx}{RX}{Receiver}
\newacronym{sa}{SA}{standalone}
\newacronym{sack}{SACK}{Selective Acknowledgment}
\newacronym{sap}{SAP}{Service Access Point}
\newacronym{sch}{SCH}{Secondary Cell Handover}
\newacronym{scoot}{SCOOT}{Split Cycle Offset Optimization Technique}
\newacronym{sdma}{SDMA}{Spatial Division Multiple Access}
\newacronym{sinr}{SINR}{Signal to Interference plus Noise Ratio}
\newacronym{sm}{SM}{Saturation Mode}
\newacronym{snr}{SNR}{Signal to Noise Ratio}
\newacronym{son}{SON}{Self-Organizing Network}
\newacronym{ss}{SS}{Synchronization Signal}
\newacronym{srs}{SRS}{Sounding Reference Signal}
\newacronym{sss}{SSS}{Secondary Synchronization Signal}
\newacronym{tb}{TB}{Transport Block}
\newacronym{tcp}{TCP}{Transmission Control Protocol}
\newacronym{tdd}{TDD}{Time Division Duplexing}
\newacronym{tdma}{TDMA}{Time Division Multiple Access}
\newacronym{tfl}{TfL}{Transport for London}
\newacronym{tm}{TM}{Transparent Mode}
\newacronym{trp}{TRP}{Transmitter Receiver Pair}
\newacronym{tti}{TTI}{Transmission Time Interval}
\newacronym{ttt}{TTT}{Time-to-Trigger}
\newacronym{tx}{TX}{Transmitter}
\newacronym{ue}{UE}{User Equipment}
\newacronym{ul}{UL}{Uplink}
\newacronym{uml}{UML}{Unified Modeling Language}
\newacronym{um}{UM}{Unacknowledged Mode}
\newacronym{utc}{UTC}{Urban Traffic Control}
\newacronym{vm}{VM}{Virtual Machine}
\newacronym{rsrq}{RSRQ}{Reference Signal Received Quality}
\newacronym{rssi}{RSSI}{Received Signal Strength Indicator}
\newacronym{crs}{CRS}{Cell Reference Signal}
\newacronym{comp}{CoMP}{Coordinated Multi-Point}
\newacronym{cran}{C-RAN}{Cloud \acrlong{ran}}
\newacronym{ca}{CA}{Carrier Aggregation}
\newacronym{cco}{CC}{Carrier Component}
\newacronym{nsa}{NSA}{Non Stand Alone}
\newacronym{embb}{eMBB}{Enhanced Mobility Broadband}
\newacronym{bsr}{BSR}{Buffer Status Report}
\newacronym{srb}{SRB}{Service Radio Bearer}
\newacronym{scm}{SCM}{Spatial Channel Model}
\newacronym{sctp}{SCTP}{Stream Control Transmission Protocol}
\newacronym{mptcp}{MPTCP}{Multi-path TCP}
\newacronym{ietf}{IETF}{Internet Engineering Task Force}
\newacronym{os}{OS}{Operating System}
\newacronym{tls}{TLS}{Transport Layer Security}
\newacronym{rfc}{RFC}{Request for Comments}
\newacronym{http}{HTTP}{HyperText Transfer Protocol}
\newacronym{nat}{NAT}{Network Address Translation}
\newacronym{api}{API}{Application Programming Interface}
\newacronym{rto}{RTO}{Retransmission Timeout}
\newacronym{psc}{PSC}{Public Safety Communication}
\newacronym{rpgm}{RPGM}{Reference Point Group Mobility}
\newacronym{ic}{IC}{Incident Command}
\newacronym{rsu}{RSU}{Road Side Unit}
\newacronym{uav}{UAV}{Unmanned Aerial Vehicle}
\newacronym{usv}{USV}{Unmanned Surface Vehicle}
\newacronym{uas}{UAS}{Unmanned Aerial System}
\newacronym{iab}{IAB}{Integrated Access and Backhaul}
\newacronym{qoe}{QoE}{Quality of Experience}
\newacronym{qos}{QoS}{Quality of Service}
\newacronym{ssim}{SSIM}{Structural Similarity Index}
\newacronym{psnr}{PSNR}{Peak Signal to Noise Ratio}
\newacronym{bs}{BS}{Base Station}
\newacronym{mu}{MU}{Multiple User}
\newacronym{ag}{AG}{Air-to-Ground}
\newacronym{af}{AF}{Array Factor}
\newacronym{ula}{ULA}{Uniform Linear Array}
\newacronym{upa}{UPA}{Uniform Planar Array}
\newacronym{lcs}{LCS}{Local Coordinate System}
\newacronym{psd}{PSD}{Power Spectral Density}
\newacronym{vq}{VQ}{vector quantization}
\newacronym{a2g}{A2G}{air-to-ground}
\newacronym{ns3}{ns-3}{Network Simulator 3}
\newacronym{mmw}{mmWave}{millimeter-wave}
\newacronym{sub6}{sub-6-GHz}{}
\newacronym{aod}{AoD}{Angles of Departure}
\newacronym{aoa}{AoA}{Angles of Arrival}
\newacronym{crt}{CRT}{Complete Report Table}

\makeatletter

\providecommand{\tabularnewline}{\\}

 \let\oldforeign@language\foreign@language
 \DeclareRobustCommand{\foreign@language}[1]{%
   \lowercase{\oldforeign@language{#1}}}

\usepackage[font=small]{caption}
\usepackage[font=footnotesize]{subcaption}

\usepackage{dblfloatfix}

\def\nb0{{\mathbf{0}}}
\def\nb1{{\mathbf{1}}}

\makeatother

\begin{document}
\history{Date of publication xxxx 00, 0000, date of current version xxxx 00, 0000.}
\doi{XX.XXXX/ACCESS.XXX.DOI}

%
\title{Network-Aware 5G Edge Computing for Object Detection: Augmenting Wearables to ``See'' More, Farther and Faster}

\author{\uppercase{Zhongzheng Yuan}\authorrefmark{1}\authorrefmark{$^*$}
\IEEEmembership{Graduate Student Member, IEEE},\\
\uppercase{Tommy Azzino}\authorrefmark{1}\authorrefmark{$^*$} \IEEEmembership{Graduate Student Member, IEEE},\\
\uppercase{Yu Hao}\authorrefmark{1},
\uppercase{Yixuan Lyu}\authorrefmark{1},
\uppercase{Haoyang Pei}\authorrefmark{1},
\uppercase{Alain Boldini}\authorrefmark{2},
\uppercase{Marco Mezzavilla}\authorrefmark{1} \IEEEmembership{Senior Member, IEEE},
\uppercase{Mahya Beheshti}\authorrefmark{2,3} \IEEEmembership{Graduate Student Member, IEEE},
\uppercase{Maurizio Porfiri}\authorrefmark{2,4} \IEEEmembership{Fellow, IEEE},
\uppercase{Todd Hudson}\authorrefmark{3,4,7},
\uppercase{William Seiple}\authorrefmark{5,6},
\uppercase{Yi Fang}\authorrefmark{1},
\uppercase{Sundeep Rangan}\authorrefmark{1} \IEEEmembership{Fellow, IEEE},
\uppercase{Yao Wang}\authorrefmark{1,4} \IEEEmembership{Fellow, IEEE},
\uppercase{J.~R.~Rizzo}\authorrefmark{2,3,4,7}}
\address[1]{Department of Electrical and Computer Engineering, Tandon School of Engineering, New York University, 
Brooklyn, NY 11201, USA}
\address[2]{Department of Mechanical and Aerospace Engineering, Tandon School of Engineering, New York University, 
Brooklyn, NY 11201, USA}
\address[3]{Department of Rehabilitation Medicine, NYU Langone Health, New York, NY 10016, USA}
\address[4]{Department of Biomedical Engineering, Tandon School of Engineering, New York University, NY 11201, USA}
\address[5]{Lighthouse Guild, New York, NY 10023, USA}
\address[6]{Department of Ophthalmology, NYU Grossman School of Medicine, 
New York, NY 10016, USA}
\address[7]{Department of Neurology, NYU Langone Health, New York, NY 10016, USA}
\address[$^*$] {Equal contribution}
\tfootnote{This work was supported in part by
the NSF grant 1952180 under the Smart and Connected Community program
as well as the industrial affiliates of
NYU WIRELESS.
In addition,
Azzino, Mezzavilla and Rangan were supported by
NSF grants 1925079, 1564142, and 1547332 
and the Semiconductor Research Corporation (SRC). 
Yuan and Wang were also supported by NSF grant 2003182. Yu Hao was also partially supported by NYUAD Institute (Research Enhancement Fund - RE132). \\
NYU, J.R. Rizzo and Todd Hudson have financial interests in related intellectual property. NYU owns a patent licensed to Tactile Navigation Tools. NYU, J.R. Rizzo, and Todd Hudson are equity holders and advisors of said company.
}

\markboth
{Yuan \headeretal: Network-Aware 5G Edge Computing for Object Detection: Augmenting Wearables to ``See'' More, Farther and Faster}
{Yuan \headeretal: Network-Aware 5G Edge Computing for Object Detection: Augmenting Wearables to ``See'' More, Farther and Faster}

\corresp{Corresponding author: Zhongzheng Yuan (e-mail: zy740@nyu.edu).}

\begin{abstract}
Advanced wearable devices are increasingly 
incorporating high-resolution multi-camera systems.
As state-of-the-art neural networks for 
processing the resulting image data are 
computationally demanding, there
has been a growing interest in leveraging
\gls{5g} wireless connectivity and
mobile edge computing for offloading
this processing closer to end-users. To assess this possibility, 
this paper presents a detailed simulation 
and evaluation of \gls{5g} wireless offloading
for object detection in the case of a powerful, new 
smart wearable called \VISION, for the Blind-and-Visually Impaired (BVI).
The current \VISION system is an instrumented
book-bag with high-resolution
cameras, vision processing, and haptic and audio
feedback. The paper considers uploading the camera 
data to a mobile edge server to perform real-time object detection and transmitting the detection results back to the wearable.  
To determine the video requirements, the paper evaluates
the impact of video bit rate and resolution on object detection
accuracy and range. A new street scene dataset with labeled objects relevant to BVI navigation is leveraged for analysis. The vision evaluation
is combined with a full-stack wireless network simulation 
to determine the distribution of throughputs and delays 
with real navigation paths and ray-tracing
from new high-resolution 3D models in an urban environment. 
For comparison, the wireless simulation considers
both a standard 4G-\gls{lte} sub-6-GHz carrier and 
high-rate \gls{5g} \gls{mmw} carrier.
The work thus provides a thorough
and detailed assessment of edge computing
for object detection with \gls{mmw} and sub-6-GHz connectivity in an application with 
both high bandwidth and low latency requirements.
\end{abstract}


\titlepgskip=-15pt
\maketitle


\glsresetall
\clearpage  
\newpage

\section{Introduction}
\label{sec:intro}

Technology in 
smart wearables is advancing rapidly
with an increasing integration of 
rich camera and sensor data
\cite{niknejad2020comprehensive,sun2017smart}.
At the same time, there has been remarkable
progress in machine vision technology for processing this visual information.
A key challenge of deploying advanced machine 
vision algorithms in the wearable setting is that 
state-of-the-art deep neural networks are
computationally demanding, particularly
for mobile devices that are limited in 
power and processing resources
for high-resolution images \cite{wu2019machine}.

Mobile edge computing combined with the massive mobile broadband capabilities of \gls{5g}
cellular wireless systems offers the possibility of offloading these computationally intensive vision processing tasks to the network edge \cite{leung2021ieee,wang2019edge}.
Importantly, \gls{5g} systems can leverage the \gls{mmw} bands 
which afford vastly greater spectrum for higher-rate and lower-latency connectivity compared to standard 4G ones \cite{rangan2014millimeter,shafi20175g,uwaechia2020comprehensive}. 
With mmWave connectivity,
a mobile device or wearable 
can upload high-resolution video data to edge servers, where
much greater computational processing can be performed while keeping resources closer to the user to reduce the overall latency.
Wireless offloading can thereby  enable support for multiple cameras for an enlarged field-of-view. Edge connectivity may also provide real-time access to data from other users, converging to new cooperative service strategies.

In this work, we study the potential of
wireless offloading of machine vision processing
for a powerful, smart wearable for the Blind-and-Visually
Impaired (BVI). The system, called \VISION (Visually Impaired Smart Service System for Spatial Intelligence and Navigation) \cite{intro-7, intro-8, intro-9, arc-1, arc-2, arc-7, arc-10, arc-11} is a human-in-the-loop, sensing-to-feedback advanced wearable that supports a host of microservices during BVI navigation, both outdoors and indoors. The current \VISION system is implemented as
an instrumented backpack; more specifically, a series of miniaturized sensors are integrated into the support straps and connected to an embedded system for computational analysis; real-time feedback is provided through a binaural bone conduction headset and an optional reconfigured waist strap turned haptic interface.  

\begin{figure*}
    \centering
    \includegraphics[width=15cm]{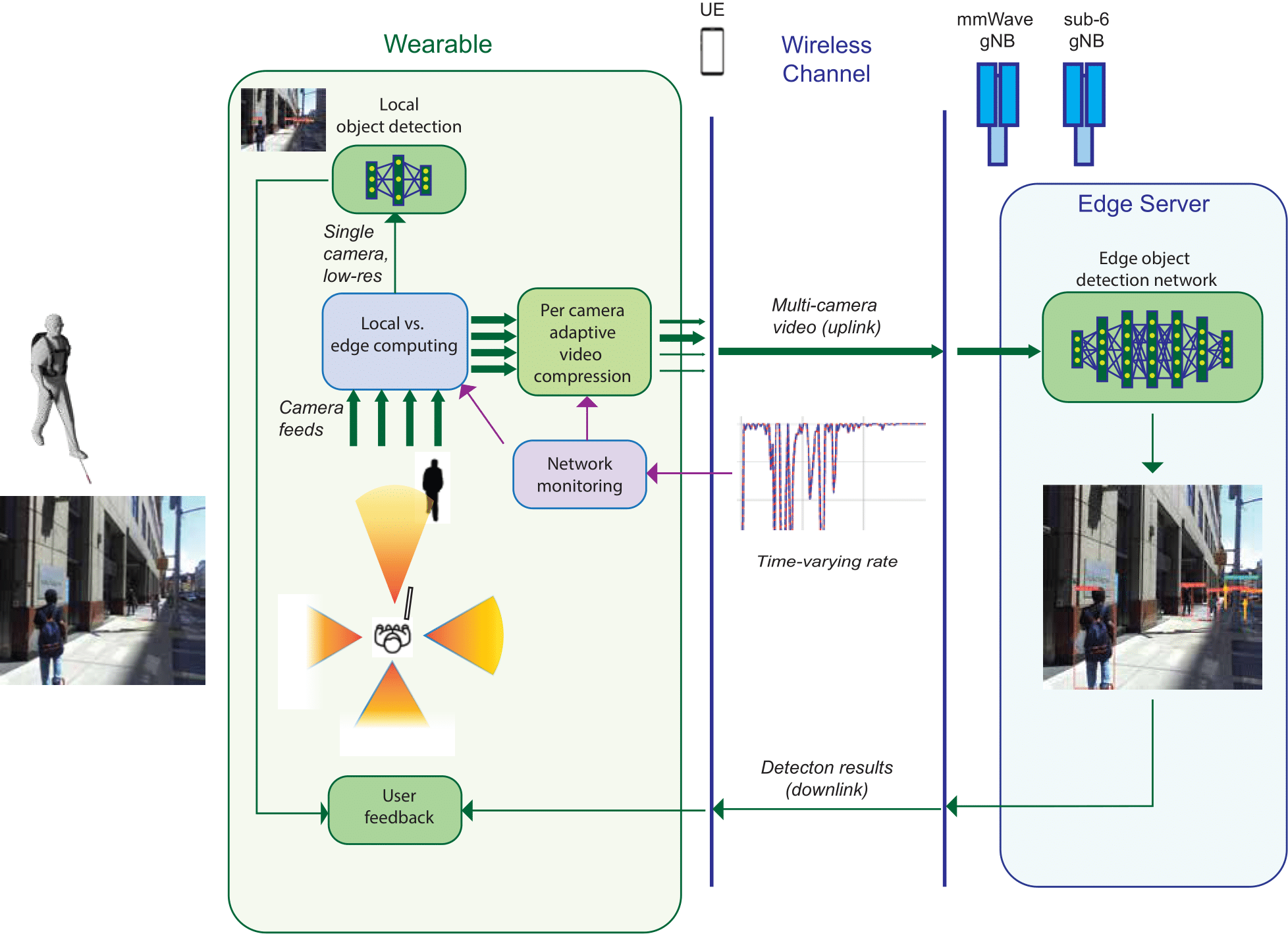}
    \caption{\textbf{Wireless offloading study:} The \VISION wearable jacket from \cite{intro-8, intro-9, arc-1, arc-2, arc-8, arc-11}
    is outfitted with multiple cameras for
    \SI{360}{\degree} view. Due to power limitations, local processing of the camera data may be limited to a single camera at low resolution. When high rate wireless
    connectivity is available, multi-camera, 
    high-resolution 
    data can be sent to an edge server where greater processing capabilities is available.  The
    detection results are then returned in the downlink. The paper assesses the feasibility of
    this approach in urban environments under
    realistic wireless channel conditions and
    deployment assumptions.}
    \label{fig:overview}
\end{figure*}

A key limitation of the current \VISION system
is that all the machine vision is performed
locally by an embedded processor in the backpack, which limits the image resolution and the frame rate at which visual computation (e.g., object detection) can be performed. Furthermore, the battery needed to enable prolonged operation adds considerably to
the backpack weight.
Here, we investigate the wireless offloading
of the vision computations to edge servers as shown in Fig.~\ref{fig:overview}.  
In the system studied, the wearable is augmented
with multiple high-resolution cameras to increase the field of view (device-wise) and enhance functionality (the current system has a single stereo camera).
When wireless connectivity is available, the camera
data will be uploaded over a cellular network to 
a mobile edge server. We analyze the system in the
case where the cellular wireless link
can include both traditional lower data rate 
carriers (e.g.,\ sub-6-GHz carriers in 4G) 
as well as higher data rate 5G 
carriers in the mmWave band.  
Since the data rate in the multi-carrier system
may be variable, we consider an adaptive video scheme where
the number of camera feeds and bit rate per
camera are adapted based on the estimated 
uplink wireless rate and delay.

The requirements for such a wireless system
are considerable and well beyond those considered in prior
vision offloading studies. For example, 
as we will see in Section~\ref{sec:video}, accurate object detection for pedestrian scenes
at reasonable distances  
can require over \SI{100}{Mbps}
if four cameras are used. Moreover, based on 
physiological markers (see Section~\ref{sec:delay_requirements}),
the total maximum end-to-end delay of the system
will likely need to be less than \SI{100}{ms}.
After removing the time for video acquisition,
compression, and inference, there is a limited
time for uplink and downlink transmission.
As described in the previous work section below, most prior applications
of edge computing for machine vision processing
with video rate adaptation 
(e.g.,~\cite{jedari2020video,li2019edge,sun2019mvideo,liu2019edge,ran2018deepdecision})
have considered relatively low-resolution,
single-camera data where the requirements
are much less strict. In these cases, 
sub-6-GHz carriers with relatively limited bandwidths
are generally sufficient.  
In contrast, we will see that the
5G mmWave bands are uniquely capable
of meeting the peak requirements for the enhanced \VISION wearable.

Yet, 5G mmWave connectivity presents considerable technical
challenges of its own when used for offloading.
Most importantly, data rates in mmWave
outdoor links are highly variable since the signals
have limited range and are strongly 
susceptible to blockage
from buildings, pedestrians, and other objects
in the environment \cite{maccartney2017rapid,slezak2018empirical}.
In addition, mmWave links are highly directional
and require continuous beam tracking to maintain
connectivity \cite{rangan2014millimeter,heath2016overview}.
This beam management and rate prediction
can cause significant additional delays
\cite{giordani2018tutorial}.
The broad goal of this paper is to provide
a detailed assessment of the feasibility of 5G mmWave
machine vision edge processing in a high data rate low-latency application.  

Our study does not include the important factor of power
consumption that arises from mmWave connectivity,
or the potential power savings from avoiding
local computation. Power analyses of mmWave
devices transceivers and beam tracking 
can be found in 
\cite{dutta2019case,skrimponis2020towards,shah2021power}, along with power measurements of commercial devices
in \cite{narayanan2021variegated}. Our focus here is on the \textit{functional} benefits
of 5G connectivity such as support for increased
number of cameras and higher resolution and 
object detection range.

Our analysis follows four main steps, each
of which bears significant new 
contributions to handle the unique nature
of the mmWave offloading system for the enhanced 
\VISION wearable:
\begin{itemize}
\item \textit{Creation of the NYU-NYC StreetScene dataset}:
First, to evaluate object detection, we curated a custom dataset,
\textit{NYU-NYC StreetScene}, of high-resolution 
    videos taken during the `Last Mile' pedestrian segment of commuting
    in NYC. The videos were manually annotated 
    with objects specific to BVI navigation.  
    A depth estimation method was developed for selected objects (standing people) so that the
    distance of the objects could also be estimated
    -- key to assessing the detection range.
    The dataset is made public and is itself
    a contribution of the work \cite{rizzolab}. 

    \item \textit{Evaluation of the impact of video resolution and bit rate on object detection:}
    Using the dataset, 
    we conducted an extensive study to evaluate the impact of 
    video resolution and bit rate on the object detection accuracy and the reliable detection range.
    As discussed in the prior work section, 
    previous analyses such as \cite{liu2019edge,ran2018deepdecision} 
    considered only low-resolution images,
    and did not explicitly study the detection range.
    Moreover, \cite{huang2017speed} did not consider the
    effect of compression.
    
    \item \textit{Wireless network evaluation:}
     We next conducted detailed, realistic wireless network simulations of end users engaged in `Last Mile' pedestrian commuting similar
     to those from which the video was captured.
    To assess the unique capabilities of 5G,
    we simulated both a 5G mmWave carrier at \SI{28}{GHz},
    and a traditional 4G \gls{lte} carrier at \SI{1.9}{GHz}.
    To accurately predict propagation at both
    frequencies, we used state-of-the-art ray tracing 
    \cite{remcom} combined with new 
    highly detailed 3D models
    acquired from GeoPipe \cite{geopipe}.
    The use of such detailed models in wireless simulation
    is the first of its kind. The channel data
    was integrated into a widely-used 
    end-to-end network simulator, ns-3 \cite{ns3}, that captures 
    blocking, beam tracking, 4G and 5G protocol functionalities as well as delays in the
    core network and edge network.
    
    \item \textit{Performance analysis and availability:} Combining the
    video and wireless analysis with inference
    times, we compared
    the performance of three scenarios:
    local processing only, offloading with LTE,
    and offloading with 5G mmWave + LTE.  
    For each option, we were able to determine key performance numbers such as the object detection accuracy, range of detection, number of cameras that 
    can be supported, and end-to-end latency.
    In addition, since the channel quality is variable,
    we determined the percentage of time that the performance 
    values can be obtained for a given scenario. We further evaluated the performance achievable with an adaptive offloading scheme, which switches between edge and local computing and varies the video resolution based on the wireless throughput.
    In summary, we completed a thorough assessment of combining state-of-the-art
    machine vision and mobile edge computing for contemporary advanced wearables.
\end{itemize}

We point out that this present work focuses on cellular wide area technologies
such as 4G and 5G since our target application is outdoor mobility. Of course,
in indoor settings and hotspots, Wireless Local Area Networks (WLANs), including high data rate versions such
as \cite{anastasi2003ieee,khorov2018tutorial}, may be
available -- see also the prior work section below
for studies in mobile edge computing with WiFi.
The study of indoor navigation with wireless
offloading with high data rate WLAN
is an interesting topic of future research.

\subsection*{Related Prior Work}
With the growing use of computationally intensive
deep learning methods for vision tasks, there has been
significant work in studying offloading of
this computation via edge computing; see, e.g., \cite{jedari2020video} for an excellent recent survey. 
However, very few consider the unique
challenges of high-resolution images transmitted over massive broadband 5G links as needed by the \VISION
system.  For example, some of the most recent
works are as follows:
\begin{itemize}
    \item Edge-AI \cite{li2019edge} studies
    edge computing for image classification.
    Since the study is on relatively limited
    data rate 4G links, the work studies
    dynamically partitioning the layers in a CNN
    for vision classification. 
    In this work, we assume the processing
    is entirely done at the edge or local device.
    \item mVideo \cite{sun2019mvideo}
    considers offloading large batches of surveillance images to the cloud for face detection.  This 
    work also only considers 4G.
    \item Hochsteetler et al. \cite{hochstetler2018embedded} studies
    inference time on a very low-power edge device 
    without access to an edge server. 
    
    \item Liu et al. \cite{liu2019edge} considered edge-assisted object detection in mobile Augmented Reality (AR)  applications. To meet the stringent delay requirement, they combine edge computing for object detection and fast local object tracking, which we adopt in our work as well. They also propose slice-based processing so that video transmission and object detection can be run sequentially over successive slices to reduce the edge computing delay. Their simulations only consider indoor WiFi connections between the VR headset and the nearby server, while we consider users walking through urban streets using 4G and 5G cellular networks. They also investigated the effect of video resolution and bit rate on the object detection accuracy and processing delay. However, they only examine resolutions  up to 720P and consider the faster R-CNN object detector, while we consider resolutions up to 2.2K and focus on the more popular YOLO detector.
    
    \item Ran et al. \cite{ran2018deepdecision} also study edge computing for AR applications. They assume the local processor (smartphone) can run either a tiny-YOLO or a  big-YOLO model, while the edge server only runs the big-YOLO model. They consider the trade-off among  decision variables including spatial resolution, frame rate, mobile power consumption, edge vs. local processing, object detection model through measurement studies and propose a measurement-driven optimization framework to determine the optimal setting for these decision variables to optimize a weighted average of the detection accuracy and frame processing rate, under the delay, bandwidth, and power constraint. However, the resolution range considered in their study is very low (only up to 480$\times$480). They also do not simulate real wireless networks.
    
      \item 
    Jiang et al.~\cite{jiang2018chameleon} propose methods to adapt frame size and frame rate as well as detection models to meet a target detection accuracy based on video content. They leverage the spatial (cross cameras) and temporal correlation of the optimal configuration to reduce the computation cost of profiling. However, this work does not consider the impact of video compression (and hence the bit rate). 
    
    \item  Huang et al.~\cite{huang2017speed}
    consider the impact  of frame size and frame rate (as well as detection model configuration) on the object detection accuracy.  This work is not evaluated in the context of edge computing and hence does not consider the effect of compression.
  
\end{itemize}

In addition to the above studies on
mobile edge computing,
there is also a large body of work on delivering 
low-latency services in 5G.  Indeed, one of 
the core design requirements of 5G are so-called
Ultra-Reliable Low Latency Communications
(URLLCs) \cite{li20185g,popovski2019wireless}
targeting air-link latencies of 1 to \SI{10}{ms}.
For the application in this paper,
the URLLC features of 5G
are critical in meeting the overall delay requirements. However, we will also see that 
several other operations contribute to the overall
delay including video framing, video encoding,
delays in the core network, and inference time.
One goal of this work can be seen as evaluating
end-to-end delays with a realistic assessment of the 
major components of the overall application.

Finally, on a more general note, the work
\cite{maier2020internet} describes an Internet of
No Things and ``seeing the invisible".  The focus 
of this paper is more limited, specifically
increasing the range and field of view via 
computational offloading.

\section{The \VISION System and Offloading Architecture} \label{sec:architecture}

\subsection{Motivation}
Immobility is a fundamental challenge for persons with BVI \cite{intro-1}. Loss of sight leads to loss of spatial cognition \cite{intro-2}. Spatial cognition can be defined as the knowledge or cognitive representation of the structure, entities/objects, and relationships within space \cite{intro-3}. The overwhelming majority of physical spaces are not visually accessible or do not allow for safe and efficient travel \cite{intro-4, intro-5}. This limits spatial cognition for the blind and leads to inefficiencies and peril during navigation. This gap requires new tools to bridge such accessibility barriers and to promote independence in daily tasks. A host of assistive technologies for citizens with BVI have been proposed. The white cane \cite{intro-8} is arguably the most widely used \cite{intro-13, intro-14} and affordable tool, but its short perceptive range limits its function as a direct extension of physical touch \cite{intro-15,intro-16,intro-17,intro-18,intro-19,intro-20}. High-tech hardware-based wearable devices have been developed to provide assistive features such as outdoor navigation \cite{chanana2017assistive,bai2019wearable,wang2017enabling}. However, they are generally either high cost or overly cumbersome \cite{wahab2011smart,elmannai2017sensor,farcy2006electronic}. In contrast, software-based solutions that run on ordinary smartphones are more affordable and accessible for BVIs. For example, Microsoft Seeing AI \cite{intro-17} and Blind Square \cite{intro-18} are widespread sensory substitution and navigation apps. However, these applications are not capable of offering advanced computer-vision-based assistive services or features due to the smartphones’ limited on-board sensing capabilities and computing power.

\begin{figure}
    \centering
    \includegraphics[trim={0 6cm 0 0 },clip,width=8cm]{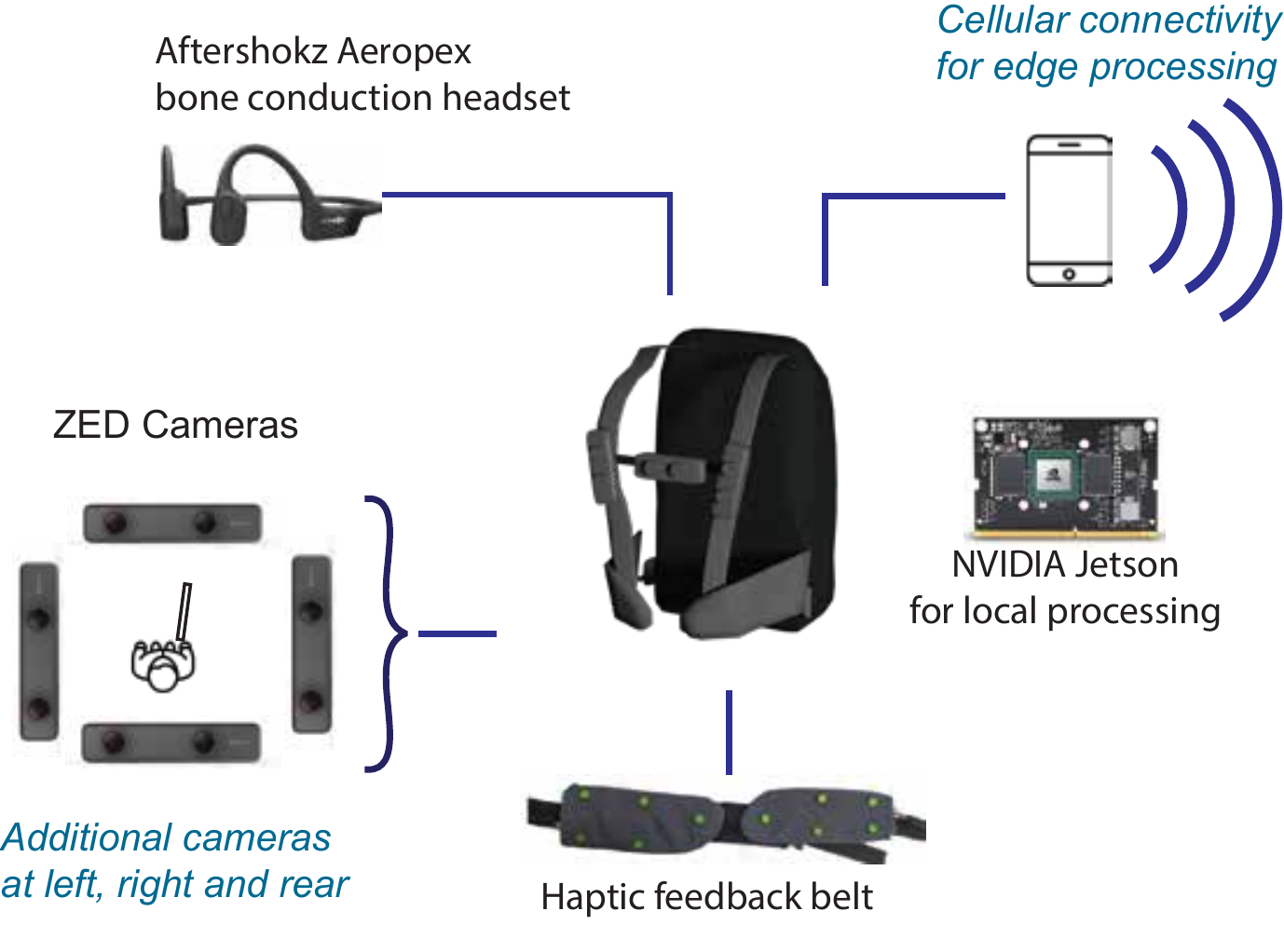}
    \caption{\textbf{\VISION} \textbf{backpack:}   The current version of the \VISION wearable jacket from \cite{intro-8, intro-9, arc-1, arc-2, arc-8, arc-11}
    is an instrumented book bag with
    a single ZED camera, NVIDIA Jetson GPU for local 
    vision processing, and haptic and audio feedback.
    In this work, we consider augmenting it
    (proposed augmentation shown in blue)
    with three additional cameras for rear, left
    and right visibility and wireless connectivity
    for edge processing.  }
    \label{fig:backpack}
\end{figure}

\subsection{The Current \VISION System and Its Limitations}

To address these challenges, we recently developed \VISION, a Visually Impaired Smart Service System for Spatial Intelligence and Navigation \cite{intro-7, intro-8, intro-9, arc-2, arc-7, intro-12}.
The system is implemented as 
a mobile sensor-to-feedback wearable device in the form of an instrumented bookbag -- See Fig.~\ref{fig:backpack}. 
This smart service system is capable of real-time scene understanding with human-in-the-loop navigation assistance, supporting both mobility and orientation \cite{arc-1, arc-8}. \VISION has four components: (1) distance and ranging/image sensors scaffolded into the shoulder straps of a backpack;  these sensors (including a stereo camera and an Inertial Measurement Unit) extract pertinent information about the environment; (2) an embedded system (micro-computer) with both computing and communication capability (inside backpack); (3) a haptic interface (waist strap) that communicates spatial information computed from the sensory data to the end-user in real time via an intuitive, torso-based, ergonomic, and personalized vibrotactile scheme; and (4) a headset that contains both binaural bone conduction speakers and a noise-cancelling microphone for oral communication \cite{intro-8, intro-9, arc-2, arc-10}. The system leverages stereo cameras as its primary sensory input and employs advanced computer vision algorithms on Nvidia Jetson processing boards. 
The goal of the embedded system is to enable continuous mapping, localization, and surveillance within a dynamically changing environment \cite{intro-8, intro-9, arc-1, arc-2, arc-8, arc-10}. 

The key limitation in the current system is that
the video processing is performed entirely
locally, which is  computationally and power intensive and limits the performance of visual analytics. 
Indeed, the wearable runs off a laptop battery with approximately \SI{66}{Wh} at \SI{0.5}{kg}  
yielding 2-3 hours of function if continuously running the vision processing.
To stay within these power limits, the wearable
uses the Jetson Xavier NX to perform object detection. With a standard YOLO model
the system is able to process 
only WVGA resolution video at a rate of 10-13 frames per second (FPS) --- See Table~\ref{tab:ComplexityVsResolution}; results are often even poorer in mobile phone applications employing similar approaches \cite{ignacio}.  
As we discuss below, the low resolution results in poor 
detection accuracy and limited range.

 
Another limitation of the current system is that it deploys only a single stereo camera providing a field of view of approximately 90 degrees horizontally and 60 degrees vertically. At about 3 meters of distance or range from the end user, a 90-degree field of view is very restrictive, leaving potentially pertinent spatial obstacles out of the perceptive capabilities of the system,  ones that may be encountered with even slight orientation shifts in forward paths. While this may be circumvented by simply using  ultra-wide angle cameras,  the geometric distortion in such cameras can degrade performance of visual analytics \cite{playout2021adaptable}. In order to address these shortcomings, we propose to embed multiple cameras in the \VISION backpack to provide omnidirectional coverage.

Although multiple cameras or 360 degrees of perception may seem superfluous for a human with no disability, a person with a disability
may benefit significantly from a system that can
provide advanced notice and anticipate danger omnidirectionally. These full-field approaches to environmental analysis are now a common practice in myriad autonomous systems, from robots to cars and drones \cite{darms2009obstacle,davidson2021fov}.


\subsection{Wireless Offloading System Studied}
Wireless offloading of vision processing 
to a mobile edge server offers two key
potential benefits:
(1) greater processing capability at the edge can enable analysis of multiple high-resolution camera streams for fast and more accurate object detection, over a wider field of view and greater range (distance); and 
(2) reducing processing on the wearable
can prolong the battery life and/or reduce the battery weight. 

To assess these potential gains, we
consider two augmentations to the backpack,
depicted in blue in Fig.~\ref{fig:backpack}.
First, we consider a version of the
wearable with four stereo cameras,
for example, to cover four sectors of 
90 degrees each.  
Second, to process multiple cameras at higher
resolution, we consider adding cellular connectivity
to an edge server with higher process capabilities.  
The system will adapt the number of camera streams to be uploaded to the edge server
and their target bit rates based on the estimated  uplink network capacity. Furthermore, the compression configuration (e.g., frame rate and spatial resolution) of each video stream will also be adjusted based on the target rates (adaptive).
The mobile edge will analyze the videos from multiple cameras using a deep learning network for object detection (and other tasks such as environmental mapping)  and send the results 
back to the wearable over the downlink;  see also
Fig.~\ref{fig:overview}.

When the wireless connection is temporarily down (e.g., due to blockage), the local processor can analyze one video stream at a lower resolution, while storing the captured high-resolution video within its on-board memory. The high-resolution video can then be opportunistically uploaded to the edge server when the user reestablishes a high-bandwidth wireless connection, to enable the mapping of various environments, perform post-hoc behavioral analysis, etc. 

\subsection{Delay Requirements}
\label{sec:delay_requirements}

For real-time pedestrian navigation, there is no generally agreed upon requirement for the tolerable total delay between the time an object appears in the environment of a pedestrian and the time it should be detected and reported to the pedestrian. In our previous work, we suggested 100 ms \cite{leigh2015neurology}. This benchmark is predicated on a physiologic marker, the high-end of the duration range for a large-amplitude saccade (fast eye movement); such an eye movement would be used by a normal-sighted pedestrian to identify a potential hazard. This stringent delay requirement enables the detection of dynamic, high-velocity objects (e.g., a suddenly appearing scooter in a pedestrian walkway). 

\section{Impact of Video Compression and Spatial Resolution on Object Detection Accuracy and Range} \label{sec:video} 

\subsection{Overview}

Due to the high bit rate of raw high-resolution video, compression is needed to stream image frames to the edge server via a throughput-constrained link.
In this section, we analyze how video compression (including reducing video resolution)
impacts object detection performance and inference time.  
The analysis will provide the bit rate and delay requirements for the wireless
uploading described in Section~\ref{sec:wireless}.

Although the augmented \VISION system will use multiple stereo cameras to enable a wider field of view and distance estimation, the analysis in this section focuses on a single monocular video. How to combine detection results from multiple views or make use of depth information in object detection is an interesting subject of
future work. 
Thus, in the wireless evaluation in the next section,
we will consider only uploading of
multiple monocular streams. 

\begin{figure}
    \centering
    \includegraphics[width=8cm]{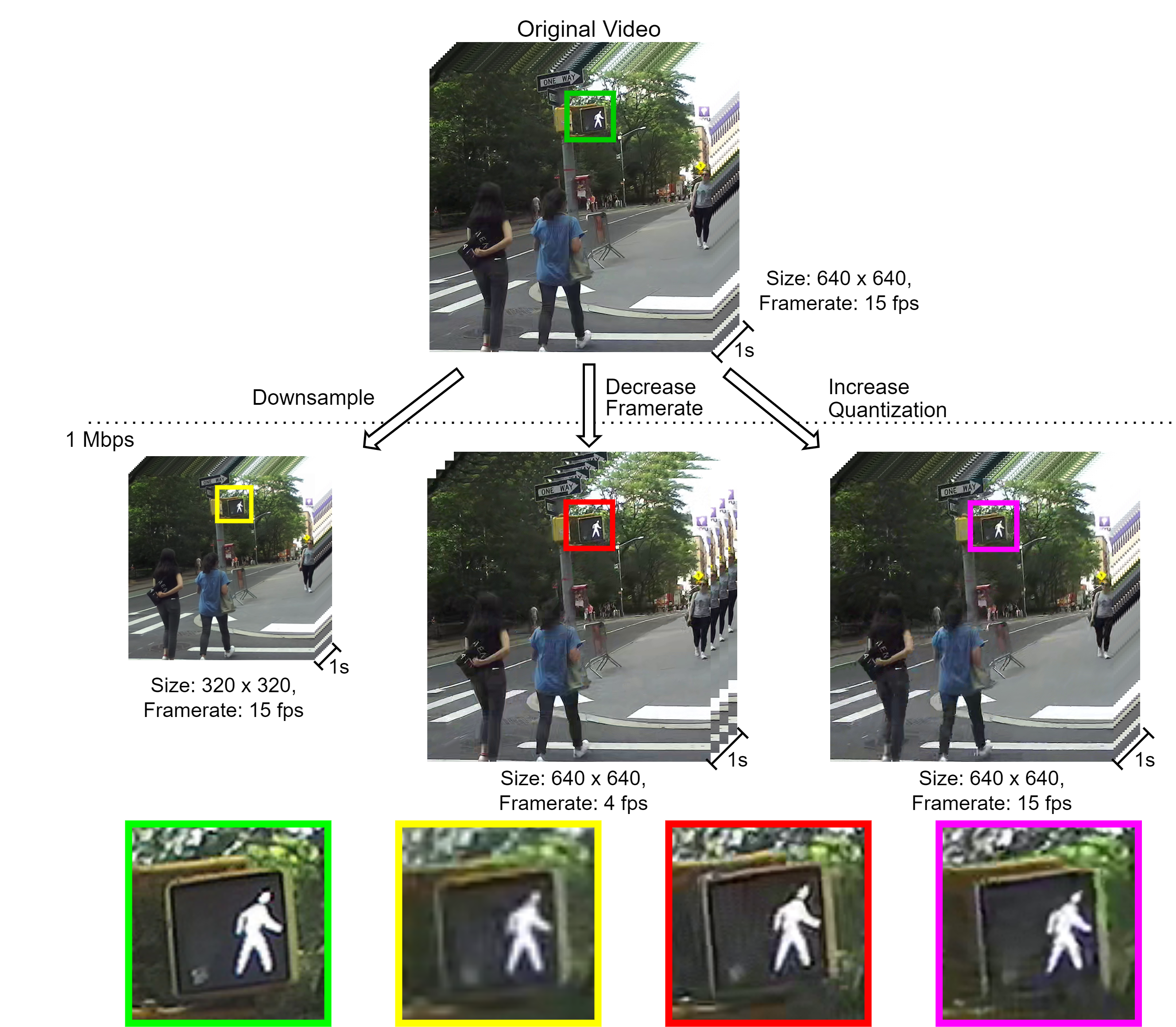}
    
    \caption{Under the same bit rate constraint, one can represent a video using different combinations of spatial, temporal, and amplitude resolutions as shown here with an example video
    compressed to \SI{1}{Mbps}. The bottom row shows a crop from each version of the video to better illustrate the differences in compression artifacts.}
    \label{fig:VideoSTAR}
\end{figure}

Given a target rate for a camera, the video can be compressed at different spatial resolutions (frame size in terms of pixels) and temporal resolutions (frame rate),  as illustrated in Fig~\ref{fig:VideoSTAR}.  With the chosen spatiotemporal resolution, the bit rate is controlled by the quantization stepsize, which controls the  amplitude resolution and affects the pixel quality.
While there has been significant work in relating 
spatial, temporal, and amplitude resolution (STAR) 
to \textit{perceptual} video quality
\cite{QSTAR,RSTAR,STARoptimization},
the effect of STAR on object detection accuracy
is less understood.  As mentioned in the Introduction,
most prior works have only studied relatively 
low-resolution images.

Here,  
we  conduct a study to systematically evaluate the impact of spatial and amplitude resolution on the object detection accuracy using a popular object detection deep learning model (YOLO \cite{YOLO}).
We leave out the consideration of the temporal resolution at this time because the YOLO model works on video frames independently. This study enables us to  determine the optimal spatial resolution for a given  bit rate, and the achievable detection accuracy under the optimal resolution at this rate. 
We will further characterize the effect of the object distance (from the camera) on the detection accuracy  under different spatial and amplitude resolutions, to provide recommendations/guidelines on the necessary spatial resolution and bit rate to meet the desired detection range for wearables that support pedestrian navigation applications. Finally, we characterize the computational cost of YOLO (including inference time) at different spatial resolutions, which provides guidance on the tolerance for the roundtrip delay with wireless offloading.

\subsection{Creation of the NYU-NYC StreetScene Data Set}
\label{sec:DataPreparation}

Currently, there are no public datasets containing high-resolution videos captured from the perspective of a typical pedestrian.  A significant effort in
this work is the creation of a new, manually-annotated
`StreetScene' video dataset for this purpose.

\paragraph*{Video collection}
To test the performance of the YOLO model for detecting objects of interest for pedestrian navigation, we recorded a set of videos  
while wearing the current 
\VISION backpack which has a single 
stereo camera on the front shoulder strap. 
The camera model is the ZED camera from StereoLabs \cite{zed2} --
 a lightweight, powerful
recording device,
ideal for wearables requiring spatial intelligence.  
A total of 9 videos were captured with the ZED
at the 2.2K spatial resolution and \SI{15}{Hz} temporal resolution, with a total video length of 43 minutes.

\paragraph*{Object annotation}
We manually annotated the bounding boxes for 15 objects of interest, listed in Table~\ref{tab:object_type}, 
along with the number of occurrences for each object. We annotated every 30th frame of each video (only left view). This annotated dataset is publicly available at \cite{rizzolab}. 
Because YOLO was trained where the `traffic light' includes both `vehicle traffic light' and `pedestrian signal',  we grouped these two separately annotated objects into the same object type when applying the YOLO model. Furthermore, because the detection performance of YOLO on `bench', `stop sign' and `dining table' is  very poor,  we only report the detection performance for detecting the remaining 11 objects.

\begin{table*}[]
\resizebox{\textwidth}{!}{
\begin{tabular}{|c|c|c|cc|c|c|c|c|c|c|c|c|c|cc|}
\hline
\cellcolor{BlueGreen!50} & \cellcolor{BlueGreen!50} & \cellcolor{BlueGreen!50}  &  \multicolumn{2}{c|}{\tblhdr{traffic light}} & \cellcolor{BlueGreen!50} &  
\cellcolor{BlueGreen!50} & \cellcolor{BlueGreen!50} & \cellcolor{BlueGreen!50}  & \cellcolor{BlueGreen!50} &
\cellcolor{BlueGreen!50}  & \cellcolor{BlueGreen!50} &
\cellcolor{BlueGreen!50} & 
  \multicolumn{3}{c|}{\tblhdr{Not reported}}
\\ \cline{4-5} \cline{14-16} 
    \multirow{-2}{*}{\tblhdr{Object}} &  \multirow{-2}{*}{\tblhdr{person}} & \multirow{-2}{*}{\tblhdr{car}} & \multicolumn{1}{c|}{
    \begin{tabular}[c]{@{}c@{}} \tblsubhdr{vehicle} \\ 
    \tblsubhdr{traffic light}\end{tabular}} & \begin{tabular}[c]{@{}c@{}}\tblsubhdr{pedestrian}\\  \tblsubhdr{signal}\end{tabular} &\cellcolor{BlueGreen!50}
\multirow{-2}{*}{\begin{tabular}[c]{@{}c@{}}\tblhdr{potted} \\ \tblhdr{plant}\end{tabular}} & \multirow{-2}{*}{\tblhdr{bicycle}} & \multirow{-2}{*}{\tblhdr{truck}} & 
\multirow{-2}{*}{\tblhdr{chair}} &
\cellcolor{BlueGreen!50}
\multirow{-2}{*}{\begin{tabular}[c]{@{}c@{}}\tblhdr{fire} \\ \tblhdr{hydrant}\end{tabular}} & 
\multirow{-2}{*}{\tblhdr{bus}} & \multirow{-2}{*}{\tblhdr{umbrella}} & \multirow{-2}{*}{\begin{tabular}[c]{@{}c@{}} \tblhdr{motor} \\ \tblhdr{cycle}\end{tabular}} & \tblsubhdr{bench}      & \multicolumn{1}{c|}{\begin{tabular}[c]{@{}c@{}}\tblsubhdr{stop} \\ \tblsubhdr{sign}\end{tabular}} & \begin{tabular}[c]{@{}c@{}} \tblsubhdr{dining} \\ \tblsubhdr{table}\end{tabular} \\ \hline
Occurrences             & 9783                    & 5442                 & \multicolumn{1}{c|}{1977}                                                             & 651                                                          & 1122                                                                     & 704                      & 459                    & 450                    & 370                                                                      & 349                  & 335                       & 162                                                                     & 247                    & \multicolumn{1}{c|}{217}                                                  & 180                                                     \\ \hline
\end{tabular}}
\caption{Object types annotated in the NYU-NYC StreetScene Dataset.}
\label{tab:object_type}
\end{table*}

\paragraph*{Video compression} 
We compressed all videos (left view only) in the StreetScene dataset using the FFmpeg software with the x265 codec \cite{FFmpeg,X265}, which follows the latest international video coding standard H.265/HEVC \cite{H265}. 
We kept the same temporal resolution, and compressed the video either at the original 2.2K spatial resolution or reduced spatial resolutions (See Table~\ref{tab:SpatialResolution}) under different quantization parameters (QPs). Default down-sampling filters (`bicubic') in FFmpeg were used for the spatial downsampling. Considering the low-delay requirement of the navigation application, we used a Group of Picture (GOP) length of 60 frames, without B-frames, i.e, each GOP starts with one I-frame, followed by 59 P-frames. 
\begin{table}[]
    \centering
    \begin{tabular}{|c|c|c|c|} \hline
       \tblhdr{Resolution}  & \tblhdr{Width} & \tblhdr{Height} & \tblhdr{Reduction factor$^*$}\\ \hline
         2.2K& 2208 & 1242 & 1x \\ \hline
         1080P & 1920 & 1080 & 0.87x \\ \hline
         720P & 1280 & 720 & 0.58x\\ \hline
         WVGA & 672 & 378 & 0.30x\\ \hline
         
    \end{tabular}
    
    \vspace{2mm}
    $^*$ Defined relative to the 2.2K resolution (same reduction in both width and height).
    \caption{Spatial resolutions considered}
    \label{tab:SpatialResolution}
\end{table}

\paragraph*{Distance estimation for standing people} 
To examine how distance affects the detection accuracy in the StreetScene dataset, which does not have accurate distance measurements
\footnote{The depth estimation from the stereo disparity in the ZED camera SDK is not very accurate and only works when the distance is within 20 meters.}, 
we developed a method to estimate the distance of standing people in our `StreetScene' dataset. 
We focus on distance estimation for
this object type since this is a relatively small object whose detection can be greatly affected by the object distance.
Given that the variation of the physical size of standing people is relatively small, the size of the box bounding a standing person is mainly determined by the distance of the person from the camera.  Based on this observation, we trained a distance estimation  model (containing a few fully connected layers) based on the bounding box width and height using the KITTI dataset \cite{geiger2012we}, which has annotated bounding boxes and distances for standing people. To account for the difference in the camera used for the KITTI data and our ZED camera, we used our ZED camera to capture a set of videos with standing people at multiple distances against a variety of backgrounds,
with the distance captured using the positional tracking system of the ZED camera. Using this dataset, we were able to learn the mapping from the distance estimated by the model trained on the KITTI data to the distance from the video captured by the ZED camera. To apply this model on the people detected in the StreetScence dataset using the YOLO model, which includes both standing and sitting people in the same object category, we looked at the distribution of the height over width ratio among the standing and sitting people in the KITTI data, and found that using a ratio threshold of 2.0 can fairly reliably separate standing people from sitting people. Therefore, we used this ratio threshold to detect standing people in the StreetScene dataset. By applying the distance estimation model to the detected bounding boxes for the standing people followed by the camera mapping, we generated the distance measurements of standing people in the StreetScene dataset. 

\subsection{Effect of Spatial and Amplitude Resolution on Object Detection Accuracy} 

\begin{figure}
    \centering
    \includegraphics[width=9cm]{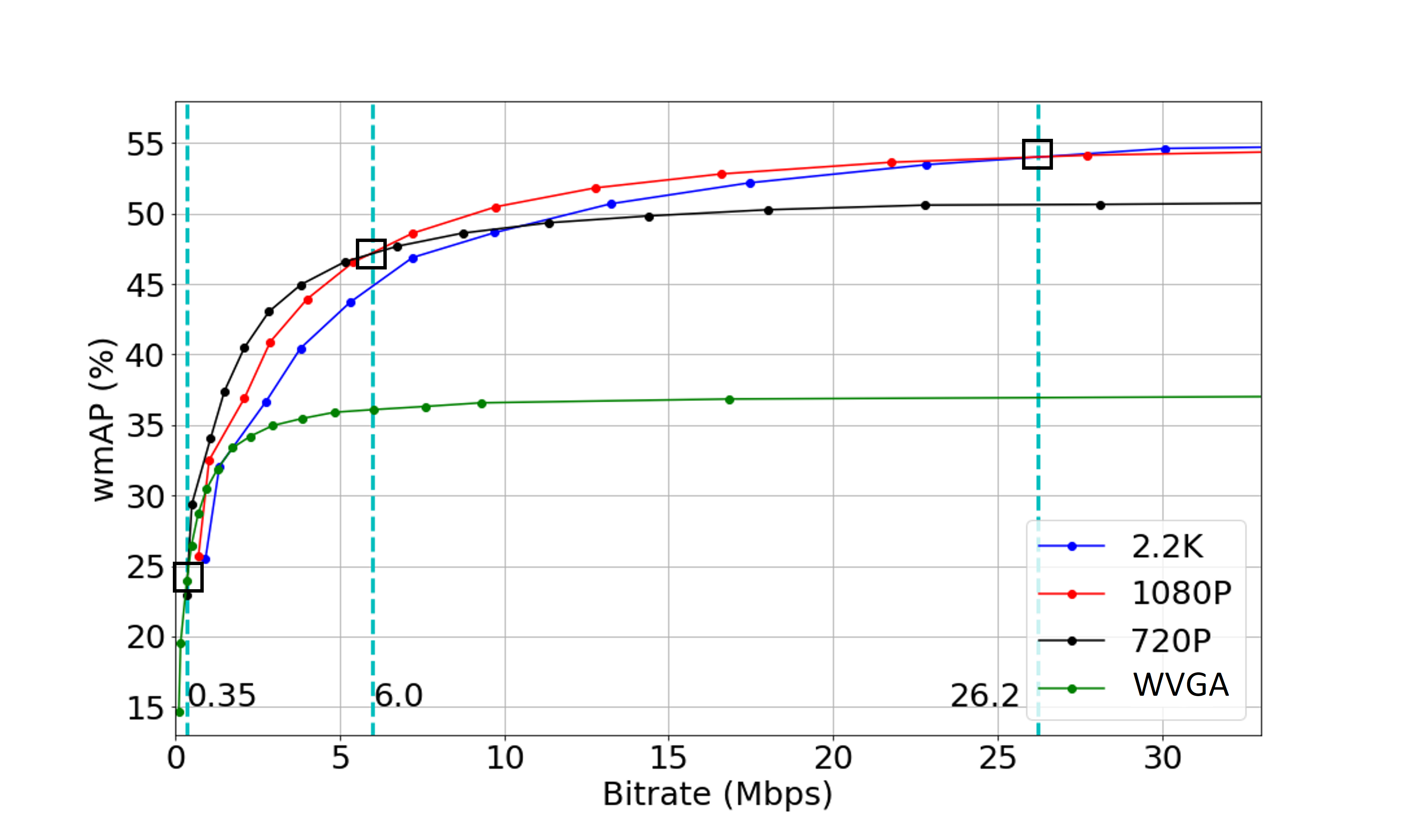}
    
    \caption{Mean detection accuracy (weighted mean AP) for 11 object types vs. bit rate for different spatial resolutions. Different points in the same curve correspond to different QPs.}
    \label{fig:wmAPvsRate}
\end{figure}

We first examine the impact of spatial and amplitude resolutions (with corresponding bit rates) on the object detection accuracy on the \textit{StreetScene} dataset, in which objects appear at varying distances. Results show that the optimal spatial resolution varies with the target bit rate (which is constrained by the network throughput). 
%
We applied a pretrained YOLO 5s model \cite{YOLO5s} on the decompressed videos in the StreetScence dataset to detect the 14 objects of interest. 
Fig.~\ref{fig:wmAPvsRate}  shows the weighted mean average precision (wmAP) over 11 objects (see Table~\ref{tab:object_type}) vs. bit rate.\footnote{Note that although the videos in the StreetScene dataset are captured and compressed  at 15 Hz, we report the equivalent bit rates for videos at 30 Hz, which is necessary to meet the real-time navigation requirement as further detailed in Sec.~\ref{sec:delay_requirements}.  This is accomplished by scaling the actual bit rates corresponding to different spatial resolutions and QPs with different scaling factors determined by a separate experiment where we compressed videos at 30 Hz and 15 Hz separately at multiple resolutions and QPs for several sample videos captured at 30 Hz.}
The weight of an object type is proportional to its occurrence frequency in the dataset. 
The figure reveals that there is an optimal spatial resolution at each bit rate that will maximize the wmAP. Specifically,  720P is best for 0.35-6.0 Mbps,  1080P for 6.0-26.2 Mbps,  2.2K for higher bit rates. However, 2.2K provides only marginal improvement over 1080P above 26.2 Mbps. We note that this could be because the YOLO model was trained mainly on low-resolution images. Although WVGA is best at a very low rate (below 0.35 Mbps), the achievable AP is too low to be usable.

At 26.2 Mbps and using 1080P resolution, the weighted mean AP is about 54\%, which is still  far from perfect. However, this relatively low detection accuracy is due to limitations of the YOLO 5s model, which was trained using uncompressed low-resolution images. Better detection models (e.g.,  models specifically trained for street scenes and/or models that are separately optimized for different resolutions) will likely further improve the detection accuracy.  It is tenable that the trend of the detection accuracy vs. rate vs. spatial resolution would be preserved for future, more powerful models.

\begin{figure}
    \centering
    \includegraphics[width=9cm]{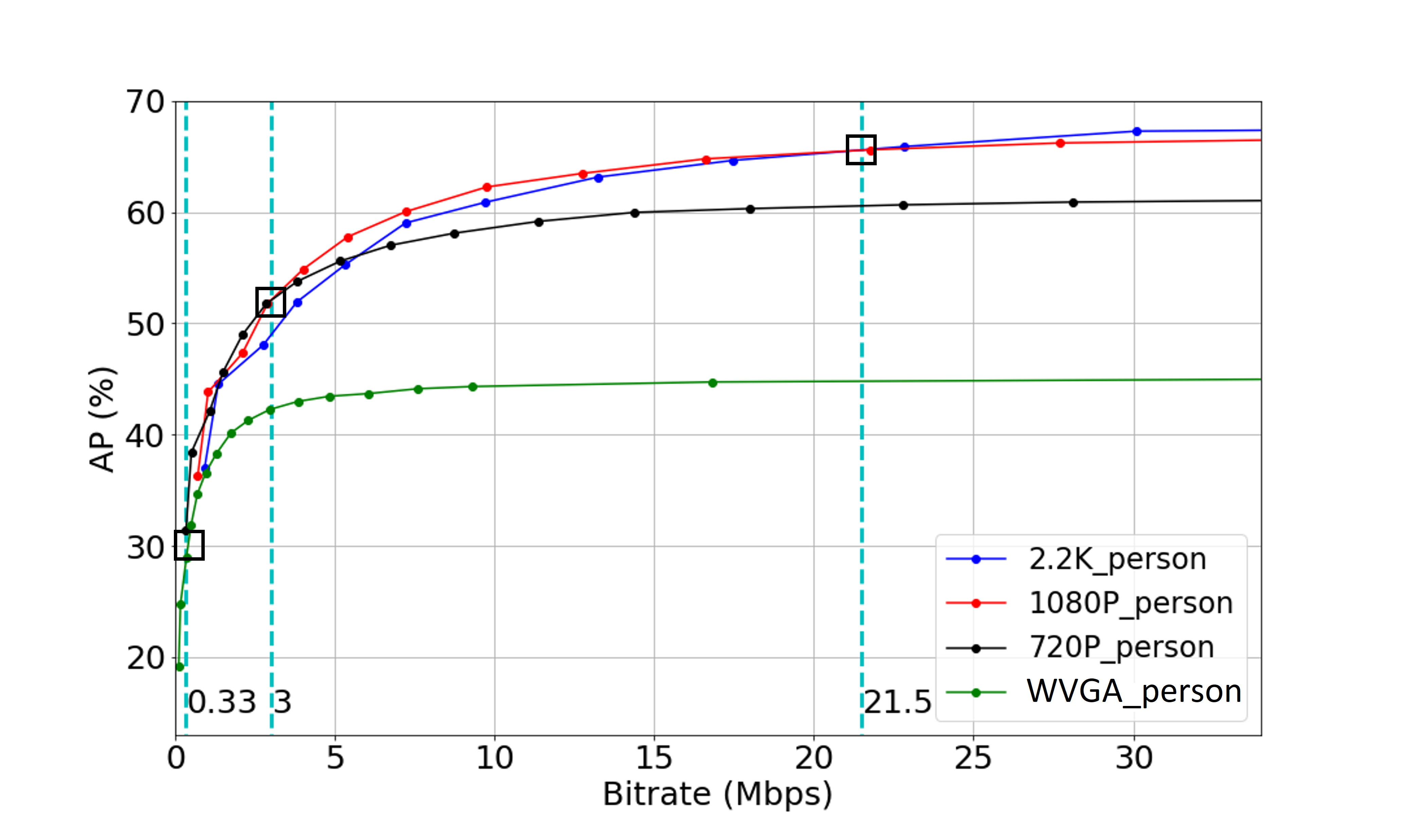}
    
    \caption{Detection accuracy (AP) for person vs. bit rate for different spatial resolutions. Different points in the same curve correspond to different QPs.}
    \label{fig:APpersonvsRate}
\end{figure}

Fig.~\ref{fig:APpersonvsRate} presents the detection result for the person category. We see a similar trend as in Fig.~\ref{fig:wmAPvsRate}, although the specific rate points where higher resolutions take over the lower resolutions are slightly different.  The AP for the person category is higher than the wmAP over 11 objects at similar bit rates, which shows that the YOLO model is more effective in detecting people than other object categories. This is consistent with the performance reported in \cite{YOLOv2}, likely because there are significantly more instances of people than other objects in the training set.  

\begin{figure*}

	\begin{subfigure}{0.49\linewidth}
		\centering
		\includegraphics[width=0.95\linewidth]{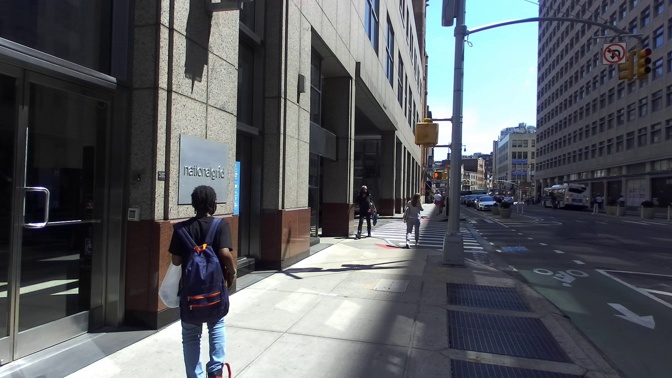}
		\caption{\textbf{Low resolution:} WVGA, QP=10, PSNR=45dB}
	
	\end{subfigure}
	\begin{subfigure}{0.49\linewidth}
		\centering
		\includegraphics[width=0.95\linewidth]{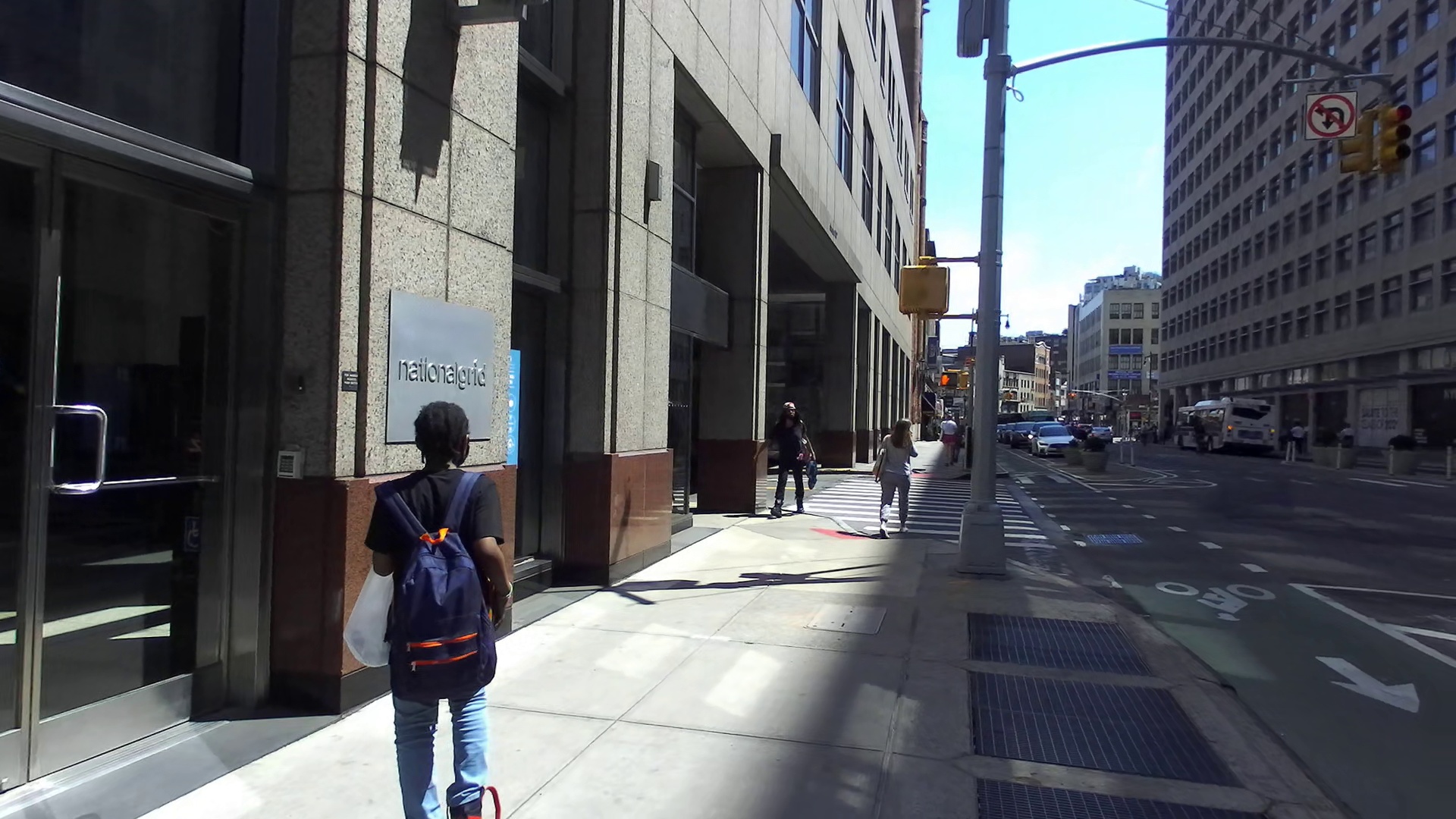}
		\caption{\textbf{High resolution:} 1080P, QP=28, PSNR=36dB}
	\end{subfigure}
	\qquad

	\begin{subfigure}{0.49\linewidth}
		\centering
		\includegraphics[width=0.95\linewidth]{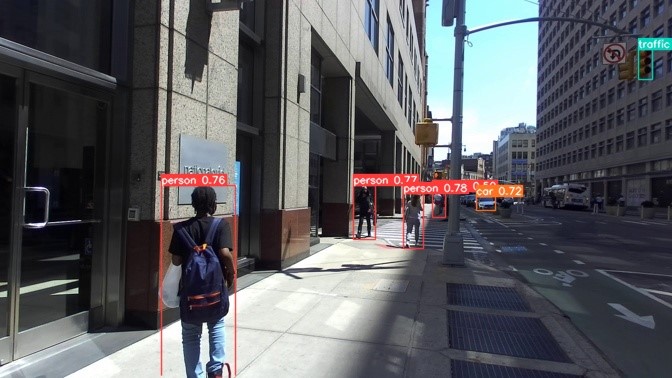}
		\caption{YOLO detection results for (a)}
	\end{subfigure}
	\begin{subfigure}{0.49\linewidth}
		\centering
		\includegraphics[width=0.95\linewidth]{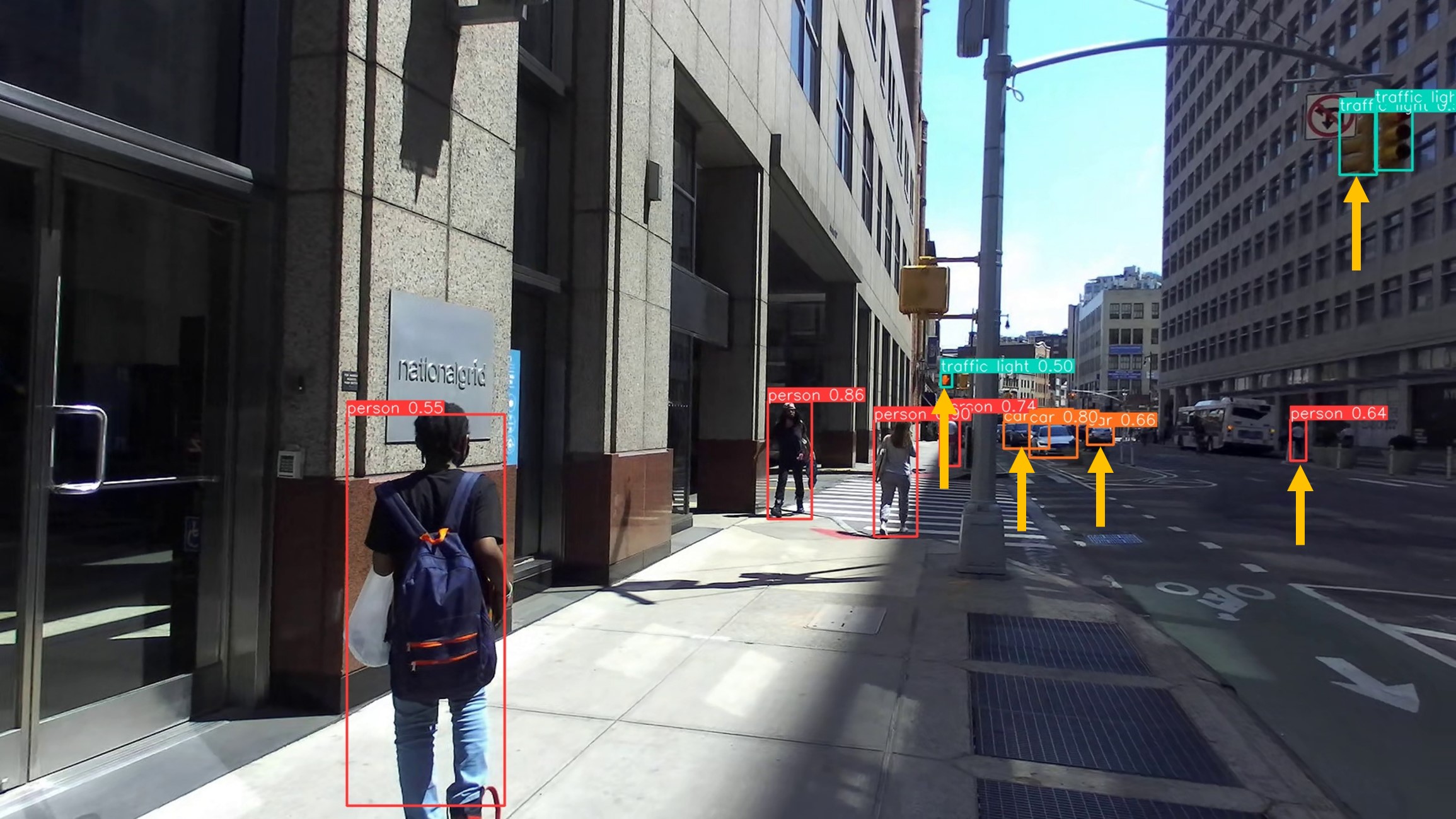}
		\caption{YOLO detection results for (b)}
	\end{subfigure}

\caption{Sample detection results from videos both compressed at 10 Mbps but using different spatial and amplitude resolutions.  Objects that are indicated by yellow arrows in (d) are missed in (c). 
Note that although the image in (a) and (c) have lower spatial resolution than those in (b) and (d), we display them at the same size for easier visual comparison. Notice that image in (a) is blurred due to the spatial upsampling. }
    \label{fig:SampleImages}
\end{figure*}

Sample frames from videos compressed to around 10 Mbps  using different settings are shown in Fig.~\ref{fig:SampleImages}(a-b).  From the outset, it is not clear  which decompressed image  will lead to improved object detection. However, from the detection results shown in Fig.~\ref{fig:SampleImages}(c-d), the YOLO model did better for  image (b), which was represented with a high spatial resolution but low amplitude resolution.

Table~\ref{tab:AccuracyVsResolution} summarizes, for each resolution, the  bit rate at which  the detection accuracy for multiple object detection (wmAP) plateaus, and the corresponding detection accuracy. We also list the AP for human detection, and the recall when the precision is 0.80. As we can see, using higher resolution video enables higher object detection accuracy, which translates to more correctly detected objects. For example, going from WVGA to 1080P, the recall for the "person" category increased from 48\% to 74\%, while keeping the false detection rate at 20\%. Therefore, by transmitting high resolution video to the edge server when the network throughput is sufficiently high, we are able to see ``more'' objects.

\begin{table}[]
    \resizebox{8.5cm}{!}{
    \centering
    \begin{tabular}{|c|c|c|c|c|} \hline
        \cellcolor{BlueGreen!50} Resolution   
        & \cellcolor{BlueGreen!50} \makecell[c]{Bitrate \\ (Mbps)} 
        & \cellcolor{BlueGreen!50} \makecell[c]{wmAP  \\ (11 obj)}  
        & \cellcolor{BlueGreen!50} \makecell[c]{AP \\ (person)} 
        & \cellcolor{BlueGreen!50} \makecell[c]{Recall \\ (person)} \\ \hline
        2.2K & 30.09 &  54.62 & 67.27 & 0.75\\ \hline
        1080P & 26.00 & 54.00 & 66.11 & 0.74\\ \hline
        720P &  18.01 & 50.27 & 60.31 & 0.68\\ \hline
        WVGA &  9.29& 36.57 & 44.32 & 0.48 \\ \hline
    \end{tabular}}
    \caption{Impact of spatial resolution on the  bit rate and the achievable detection accuracy. Recall for the person category is evaluated when the precision is 80\%. }
    \label{tab:AccuracyVsResolution}
\end{table}

\subsection{Effect of Spatial Resolution on Detection Accuracy at Different Distances}

The performance measures reported so far are aggregated results for objects appearing at varying distances from the camera. Generally, detecting a faraway object is harder than a nearby object. On the other hand, being able to detect an object while it is still far away provides more time for navigation planning. Therefore, it is important to understand how the detection accuracy degrades as the distance increases and what is the maximum distance when an object can be detected reliably. How does the distance affect the detection also depends on the physical size of the objects. We show such results for the detection of standing  pedestrians as a case study, wherein we use the algorithm described in Sec.~\ref{sec:DataPreparation} to detect standing pedestrians and furthermore estimate the distance of the detected person(s) from the bounding box size(s).

\begin{figure}
    \centering
    \includegraphics[width=0.95\linewidth]{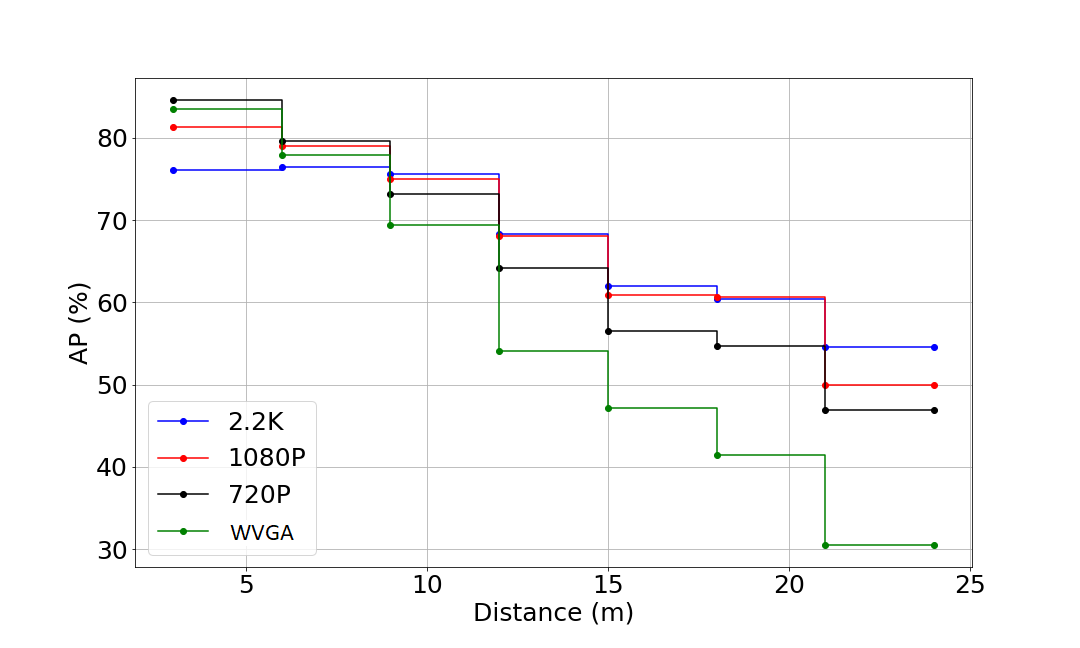}
    \caption{Detection accuracy (in AP)  vs. distance for standing people under different resolutions. The results are obtained when videos at different resolutions are compressed to the bit rates listed in Table~\ref{tab:ComplexityVsResolution}.  Note that the results  would be similar even if the videos were uncompressed because the average detection accuracy for each particular resolution already plateaued at its corresponding bit rate,  as shown in Fig.~\ref{fig:APpersonvsRate}.}
    \label{fig:standing_people_AP}
\end{figure}

We quantized the distances to several bins and determined the AP within each bin.
Fig.~\ref{fig:standing_people_AP} illustrates how the detection accuracy drops as the object distance increases under different spatial resolutions. When the distance is very close, YOLO performs very well even at the WVGA resolution, but the accuracy drops quickly as the distance increases at this low resolution. As expected, higher spatial resolutions enjoy a slower decay rate. The  2.2K resolution leads to significantly better detection than 1080P only when the distance is  greater than 21 m.  Note that the 2.2K resolution did worse than other lower resolutions in the short distance range  in this study. This is likely because the YOLO 5s model was trained on low-resolution video (close to WVGA). People within a short distance occupy a very large area in the 2.2K image, requiring bounding box sizes that rarely occur in the training data.

\begin{figure}
    \centering
    \includegraphics[width=0.95\linewidth]{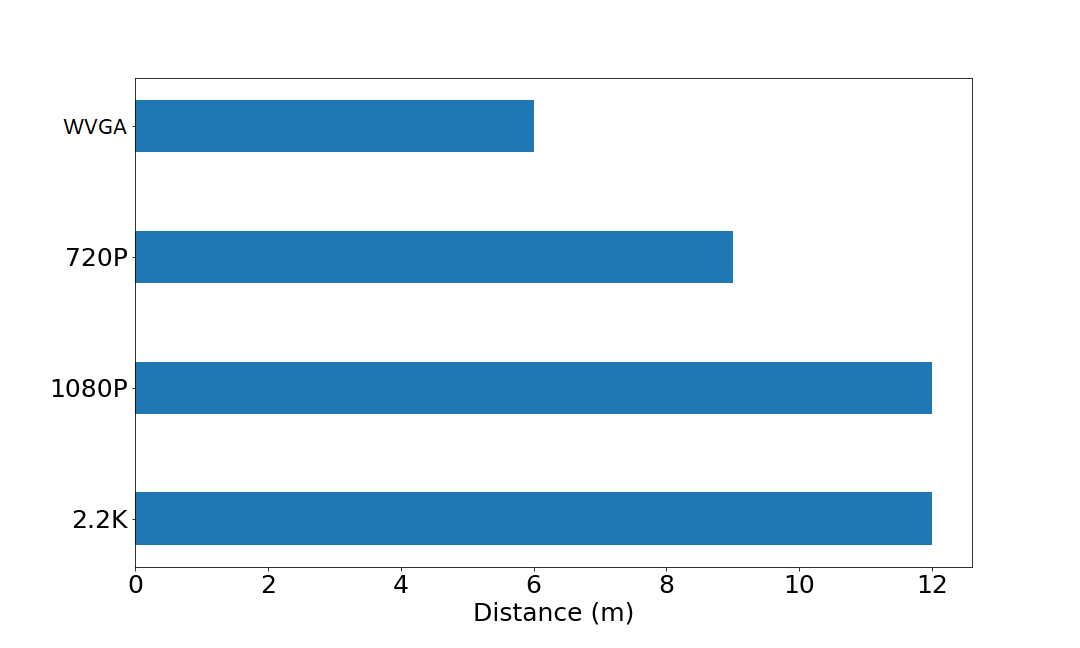}
    \caption{Detection range vs. spatial resolution for standing people. The detection range is defined as the  maximum distance at which  recall $\geq 0.9$ and precision  $\geq$ 0.83. Videos' bit rates are listed in Table III.}
    \label{fig:standing_people_range}
\end{figure}

Fig.~\ref{fig:standing_people_range} shows the detection ranges for different spatial resolutions. Here, we see clearly that going from WVGA to 1080P, we are able to extend the detection range from about 6 m to 12 m. This study shows that  we should use at least 1080P video to be able to reliably detect people  at a distance important for navigation planning. 

\subsection{Computational Complexity vs.\ Spatial Resolution}

Table~\ref{tab:ComplexityVsResolution} summarizes the computation complexity (measured by the FLOP count), the inference time per video frame, and corresponding speed (frame/sec or fps) on the embedded processor in our \VISION backpack (Jetson Xavier NX running at 15 Watts, using a GPU at 1.1 GHz), the inference time per video frame and speed using an edge server equipped with an RTX 8000 GPU, of  the YOLO 5s model, for videos at different spatial resolutions. 
As will be explained in Sec.~\ref{sec:evaluation}, to meet the total delay requirement for real-time navigation, local processing should be completed within 67 ms, which is barely possible with the WVGA video, severely limiting the achievable object detection accuracy and detection range (cf. Table~\ref{tab:AccuracyVsResolution}, Fig.~\ref{fig:standing_people_range}). On the other hand, offloading the computation to the edge server allows us to process the 1080P video and consequently significantly increase the detection performance, while still meeting the delay constraint.

\begin{table}[]
    \resizebox{8.5cm}{!}{
    \centering
    \begin{tabular}{|c|c|c|c|c|c|} \hline
        \cellcolor{BlueGreen!50} Resolution & \cellcolor{BlueGreen!50} GFLOP & \cellcolor{BlueGreen!50} \makecell[c]{Local  \\ Inference \\Time (ms)\\} &
        \cellcolor{BlueGreen!50} \makecell[c]{Local  \\ Inference \\Speed (fps)\\} &\cellcolor{BlueGreen!50} \makecell[c]{Server \\ Inference \\Time (ms)}   
        &\cellcolor{BlueGreen!50} \makecell[c]{Server \\ Inference \\Speed (fps)}  
        \\ \hline
        2.2K & 57.22 & 232.02 & 4.31 & 23.4 & 42.7 \\ \hline
        1080P & 43.38 & 178.25 & 5.6  &18.7 &53.5 \\ \hline
        720P & 19.56 & 95.69 & 10.5 & 10.4 &96.2\\ \hline
        WVGA & 5.36 & 75.02 & 13.3 & 5.1 & 196.1 \\ \hline
    \end{tabular}}
    \caption{Impact of spatial resolution on the detection model complexity, running time on local processor and edge server, respectively.  The Jetson Xavier NX is used as the local processor, while the server uses an RTX 8000 GPU.  }
    \label{tab:ComplexityVsResolution}
\end{table}


\section{Wireless Evaluation} \label{sec:wireless}


Having analyzed the bit rate and latency 
requirements for video processing, we now
simulate the wireless network to determine
what percentage of time these requirements can be met.


\subsection{User Route and Ray Tracing} \label{sec:raytracing}



\begin{figure}%
    \centering
    \includegraphics[width=0.95\linewidth]{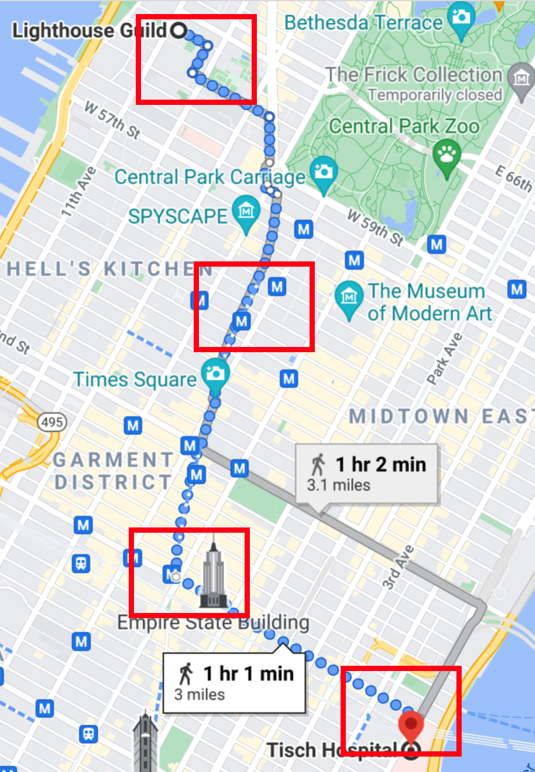}%
    \caption{Main walking route from which we extracted four specific sites represented by the red rectangles.  Top to bottom, the sites are \textit{Lighthouse Guild}, \textit{Midtown}, \textit{Herald Square}, and \textit{NYU Langone}}
    \label{fig:main_route}
\end{figure}

We simulate a hypothetical end-user commuting through the streets of Manhattan -- a challenging environment from a wireless perspective,
due to the tall building blockage.
In particular, we identified a walking route starting at Lighthouse Guild (a healthcare and research center that assists persons with BVI) and finishing at the NYU Tisch Hospital in Kips Bay, as depicted in Fig.~\ref{fig:main_route}. The environment of this route is similar to where the NYU-NYC StreetScene dataset described in Section~\ref{sec:video} was collected. 
Along this route, we selected four sites (red rectangles of Fig.~\ref{fig:main_route}) to perform a realistic full stack \gls{5g} simulation. This simulation involved two main steps: 1) generation of ray-tracing data of the wireless environment from the realistic 3D layout of the city at these sites, and 2) end-to-end simulation using \gls{ns3} and the ray-tracing information obtained in the previous step. Here, we describe the generation of ray-tracing data with the figures for the top site in Fig.~\ref{fig:main_route} as an example.

\begin{figure}
    \centering
    \includegraphics[width=0.95\linewidth]{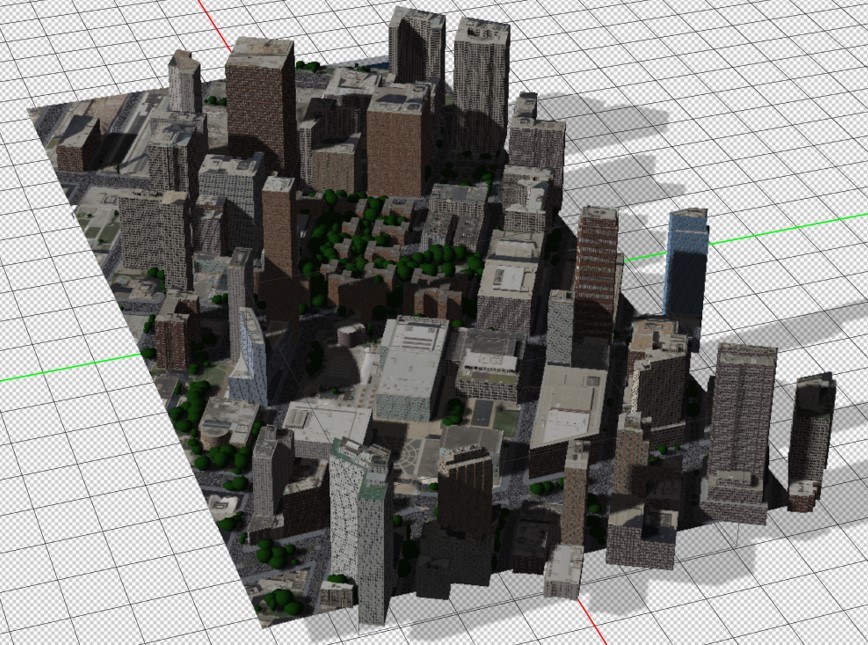}%
    \caption{3D map of the top site, \textit{Lighthouse Guild}, in the route
    in Fig.~\ref{fig:main_route}, with material information for each building and vegetation
    provided courtesy of GeoPipe \cite{geopipe}. 
     }%
    \label{fig:3d_map}%
\end{figure}

In wireless communications, ray-tracing involves the calculation of the paths that electromagnetic waves, represented as rays, follow while propagating in a known 3D environment according to a set of transmitting and receiving positions. For each combination of transmitting and receiving locations, the output of a ray-tracing simulation consists of a set of propagation information for each ray: path loss, propagation time, phase offset, \gls{aod}, and \gls{aoa}. We used Remcom's \textit{Wireless InSite} \cite{remcom} software to generate ray-tracing data in our work.  
This software has been successfully used in a number
of other studies  \cite{XiaRanMez2020,Alkhateeb2019,khawaja2017uav}.
Accurate ray-tracing of a 3D scenario at \gls{mmw} and sub-6-GHz frequencies requires precise information of buildings materials, since each material affects the propagation of the electromagnetic wave differently at each frequency range. In addition, details of the vegetation in a certain area are essential for accurate ray-tracing at \gls{mmw} frequencies. To capture all this information, we imported in Remcom the 3D layout of the City at each of the 4 sites, as provided by Geopipe \cite{geopipe}.
Fig.~\ref{fig:3d_map} depicts the 3D layout with building materials and vegetation at the top site of Fig.~\ref{fig:main_route}. 
This data from GeoPipe provides 
one of the most accurate models for \gls{mmw}
ray tracing. In particular, the models contain
small building features which are known to influence
mmWave propagation significantly \cite{rappaport2015millimeter}.
For each material (i.e., concrete, wood, glass, etc.) and frequency range, Remcom uses different parameters for the reflection and diffraction coefficients, as well as different propagation properties.
We performed the simulation at two frequencies:
\SI{1.9}{GHz} for a typical sub-6 GHz 4G LTE carrier,
and \SI{28}{GHz} for a typical 5G \gls{mmw} carrier.

Moreover, we considered a real placement of the base stations in each of the four sites as provided by \cite{opendata}. This database contains information regarding the actual foreseen placement of \gls{5g} \gls{mmw} \glspl{bs} in New York City. We assumed that the LTE 4G and 5G mmWave base stations are 
co-located, meaning that, at each
cell site, base station equipment is available for  both frequencies.  This co-location is common
since, once the operator has secured a site, it
generally utilizes it maximally.  Note that 
in 3GPP terminology,
a 4G base station is called eNB (evolved Node B),
and a 5G base station is called gNB (next generation Node B). 


For ray tracing, each base station site  represents a transmitting location, 
whereas the receiver positions are placed one meter apart along the route taken by the hypothetical user at each specific site (more details in \ref{sec:wresults}).  The ray tracing then 
provides the channel from each \gls{bs} to each
position along the route at both frequencies.
Note that due to symmetry, the large scale
channel parameters are identical in the uplink
and downlink.  Hence, we can use the estimated
channel in both directions.

\subsection{Network Simulation}
    
The second step in the evaluation of wireless offloading 
involves the end-to-end (full stack) network simulation using \gls{ns3}. \Gls{ns3} is an open-source, discrete event network simulator which affords end-to-end simulations (i.e., simulations that 
model the entire network stack from the physical layer up to the application layer) with support for user mobility and traffic modeling, among other features. To compare the performance of \gls{mmw} connectivity with a standard sub-6 GHz system,
we run two simulations:
\begin{itemize}
    \item An 4G LTE system at \SI{1.9}{GHz} 
    with \SI{40}{MHz} downlink and \SI{40}{MHz}
    uplink total bandwidth; and
    \item A 5G mmWave system at \SI{28}{GHz}
    with \SI{400}{MHz} total bandwidth
    that is Time-Division Duplexed (TDD) for the
    uplink and downlink.
\end{itemize}
For both the 4G and 5G systems, the deployment
would likely be on multiple carriers,
as is common today. For example, 
the LTE system could be two standard carriers
of \SI{20}{MHz} each and the 5G mmWave
system with four carriers of \SI{100}{GHz} each.
The parameter values for both systems are shown
in Table~\ref{tab:params} and are representative
of typical 4G and 5G simulation studies,
see, e.g., \cite{akdeniz2014millimeter,3gpp-channel,shafi20175g}.
In addition, details for the 4G-LTE \gls{ns3} module can be found at \cite{lte-ns3}.
To account for loading, we assume an individual UE
obtains a fraction $0.25$ of the total 
bandwidth, which would represent a moderate loading
level relative to standard evaluation methodologies
\cite{3gpp-channel}.
Hence, we simulate the 4G user as operating
in a system with $10+10$\, \si{MHz} bandwidth
and the 5G user as operating in a system
with \SI{100}{MHz} total bandwidth. Morevoer, the \gls{5g} system operating at mmWave frequencies uses Numerology 2 and the TDD configuration allows symbols to be \textit{flexible}, meaning that each symbol can be used for either uplink or downlink traffic.

In both the 4G and 5G cases, we model the wearable as a \gls{ue} traversing the
path described above. The channels from the ray tracing in Section \ref{sec:raytracing} are imported into the \gls{ns3} simulator.

Since ray tracing captures only the 
buildings and foliage, it does not capture blockage
from objects such as humans and vehicles in the
environment. As discussed in the Introduction,
modeling blockage is critical to accurately
assess mmWave coverage \cite{maccartney2017rapid,slezak2018empirical}.
To model this additional blockage,
we employ the \textit{Blockage Model A} in \cite{3gpp-channel} which is integrated into the \gls{ns3} simulator \cite{ns3}.
This model adopts a stochastic approach for capturing human and vehicular blocking. In particular, multiple 2D angular blocking regions, in terms of azimuth and elevation angular spreads, are generated around the \gls{ue}. One blocking region, denoted as self-blocking region, captures the effect of human body blocking, whereas $K_{\rm NSB}$ non-self-blocking regions with random sizes are used to model other sources of blockage ($K_{\rm NSB}$ can be changed to increase/decrease the density of blockers). Once the blocking regions are computed, each cluster (or ray, in the case of our ray-tracing data) is attenuated accordingly, based on the angular spreads and position of each blocking component. The parameter $T_{\rm NSB}$ denotes the time interval at which new blockers are randomly generated.


\begin{table}[t!]
\caption{Wireless network simulation parameters}
\label{tab:params}
\footnotesize
\centering
\begin{tabular}{|>{\RaggedRight}m{3.3cm}|>{\RaggedRight}m{1.6cm}|>{\RaggedRight}m{1.6cm}| }
\hline 
\cellcolor{BlueGreen!50} & \multicolumn{2}{c|}{\tblhdr{Value}} \\ \cline{2-3} 
\multirow{-2}{*}{\tblhdr{Parameter}} &  \tblsubhdr{4G LTE} & 
\tblsubhdr{5G mmWave} \\ \hline
Carrier frequency  & \SI{1.9}{GHz} 
& \SI{28}{GHz} \\ \hline
Total bandwidth (MHz) & 40+40 FDD & 400 TDD  \\ \hline 
Fraction of bandwidth available
to UE due to loading & 0.25 & 0.25 \\ \hline
Bandwidth to UE (MHz) & 10+10 FDD & 100 TDD  \\ \hline 
UE array & Single & $4 \times 4$ \\ \hline
BS array & Single & $8 \times 8$ \\ \hline
UE transmit power (dBm) & 25 & 25 \\ \hline
DL transmit power (dBm) & 30 & 30 \\ \hline
UL noise figure (dB) & 5 & 5 \\ \hline
DL noise figure (dB) & 5 & 5 \\ \hline
Number of HARQ processes, $N_{\rm HARQ}$ & 8 & 20 \\ \hline
Non-self-blocking components, $K_{\rm NSB}$ & 
\multicolumn{2}{c|}{40}  \\ \hline
Blockage update period, $T_{\rm BLK}$ & 
\multicolumn{2}{c|}{\SI{100}{ms}}  \\ \hline
TCP packet size & \multicolumn{2}{c|}{\SI{1024}{B}}
\\ \hline
Core network delay $D_{\rm core}$ & 
\multicolumn{2}{c|}{\SI{5}{ms}} \\ \hline
\end{tabular}
\end{table}

Given the blocked channels, the \gls{ns3} 
simulator then models the full stack communication.
At the Physical (PHY) and \gls{mac} layers, the modeling 
includes beam tracking, \gls{cqi} reports, rate prediction, \gls{harq}, and scheduling. The total number of \gls{harq} processes for both systems is specified in Table~\ref{tab:params}.
At the higher layers, the simulator models all the \gls{rlc}
segmentation and buffering, \gls{rrc} signaling, and handovers.
In our simulations, we use the Acknowledged Mode (AM) for the \gls{rlc} layer for both 4G and 5G systems. 

An important parameter in the network configuration
is the location of the edge server.  In 
cellular systems, data in the uplink
traverses a path:  UE (wearable) $\rightarrow$
 base station (4G eNB or 5G gNB) $\rightarrow$
 core network $\rightarrow$ server in the public
 Internet.
The downlink follows the reverse path.
In conventional deployments, operators have 
relatively few gateway points from the core network
to servers in the public Internet.
The data may thus need to traverse a long path 
in the core network to the closest gateway
resulting in high delay from
the base station to the server ---  see, for example, measurements in commercial networks in \cite{narayanan2021variegated}.
Mobile edge computing reduces the core network
delay by placing the edge servers much closer
to the base station \cite{hu2015mobile}.
In this study, we will assume that the
one-way delay from the base station to the 
edge server is $D_{\rm core} = 5$\, \si{ms}.
As we will see in the delay analysis below,
this lower delay will be critical to meet
the strict delay requirements for BVI navigation.


\subsection{Traffic Modeling}
Ideally, in the uplink,
we would model the adaptive 
multi-camera
video encoding 
application in the \gls{ns3} simulator. This analysis would
then be specific to the video adaption algorithm
used. To provide a more general and simpler
analysis, we instead model the uplink video 
data as a single Transmission Control Protocol (TCP) stream with a full buffer up to 
a maximum data rate of \SI{120}{Mbps}.
This maximum data rate is sufficient to support
four cameras at \SI{30}{Mbps} each.  
Since TCP has congestion control, it will
automatically adjust the sender rate to 
the available uplink link capacity.
As a simplification, we  
assume that the video encoding
can be adapted to be exactly the same as the TCP rate.
Hence, the full buffer 
TCP rate at any time can be regarded
as an approximation of the actual video
rate.

For the downlink, data from 
the edge server to the wearable is used to carry
the object detection results.
We model this downlink traffic as a constant
bit rate application at 30 packets per seconds,
corresponding to the expected video frame rate.
We assume that the total rate is \SI{1}{Mbps},
which is ample to specify a large number of detected objects, including their bounding boxes and probabilities belonging to different object classes.

For low-latency streaming
applications, one should not use 
TCP as a transport
protocol. For example, one can use User Datagram Protocol (UDP) or Real-time Transport Protocol (RTP) that are designed for real-time applications \cite{schulzrinne1996rtp}. Here, we use TCP only to simulate the available link rate, since TCP’s congestion control automatically adjusts to the link rate and therefore can be regarded as a proxy for the video rate adaptation that would need to be incorporated on top of any real-time transport protocol such as RTP.

\subsection{Simulation Results}
\label{sec:wresults}

\begin{figure}%
    \centering
    \includegraphics[width=0.9\linewidth]{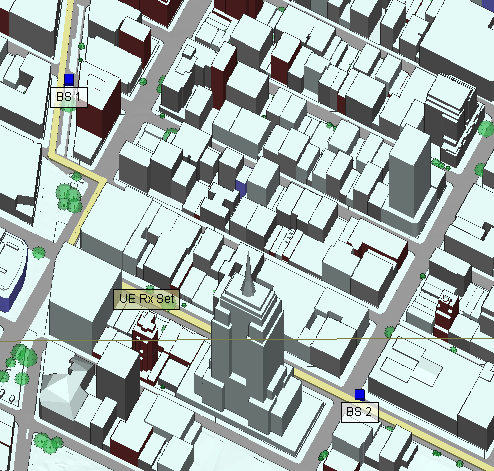}%
    \caption{\Gls{ue} route for the \textit{Herald Square} displayed along with the 3D model used in Wireless InSite ray tracing tool
    \cite{remcom}. The \gls{ue}'s RX positions are highlighted by the yellow line. \gls{bs} 1 and \gls{bs} 2 provide both 4G-\gls{lte} and \gls{5g} \gls{mmw} connectivity.}%
    \label{fig:oneroute}%
\end{figure}

\begin{figure*}
	\begin{subfigure}{0.5\linewidth}
		\centering
		\includegraphics[width=0.95\linewidth]{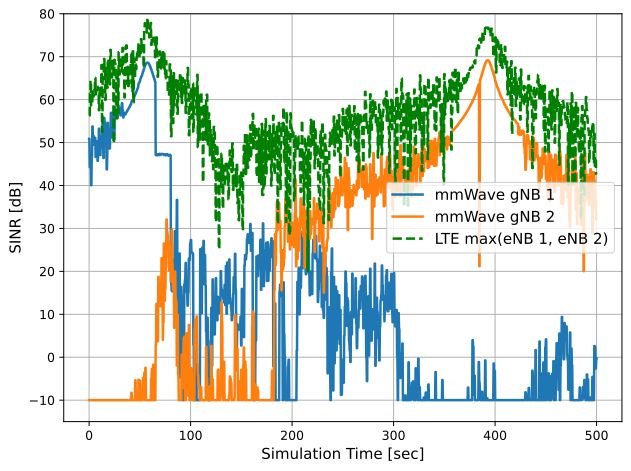}
		\caption{\Gls{sinr} over time in dB.}
	\end{subfigure}
	\begin{subfigure}{0.5\linewidth}
		\centering
		\includegraphics[width=0.95\linewidth]{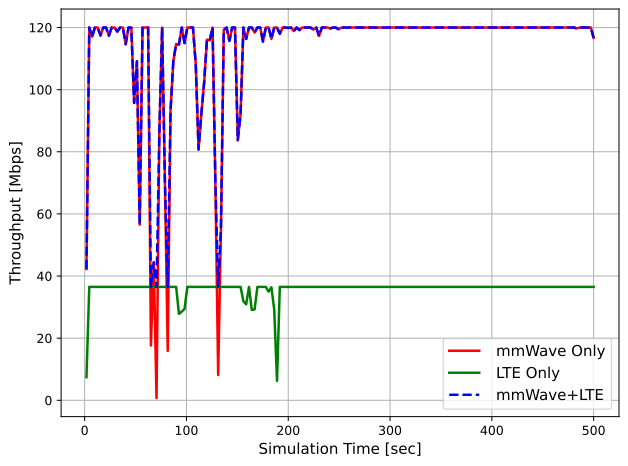}
		\caption{End-to-end throughput over time in Mbps.}
	\end{subfigure}
    \caption{UE \gls{sinr} and end-to-end throughput over time for the \textit{Herald Square} site.}
    \label{fig:herald}
\end{figure*}


As mentioned above, we performed end-to-end wireless simulation of an end user walking through four sites in Manhattan, near \textit{Lighthouse Guild}, \textit{Midtown}, \textit{Herald Square}, and \textit{NYU Langone} 
as shown in Fig.~\ref{fig:main_route}.
As one example, 
Fig.~\ref{fig:oneroute} depicts the route of the user (\gls{ue}) for the \textit{Herald Square} site. The \gls{ue} starts walking from the north-west corner in the figure and moves south-east across 34$^{\rm th}$ Street (yellow line in the picture). In this site, two base stations  are present, both
with 4G-\gls{lte} and 5G. Figure~\ref{fig:herald}(a) shows the \gls{sinr} over time for this scenario for the two \gls{mmw} base stations (gNB 1 and gNB 2), as well 
as the maximum SINR for the two LTE cells.
We see that, in this case, the LTE SINR
is continuously high (mostly $>$\SI{40}{dB})
due to the favorable propagation of 
the lower frequency (\SI{1.9}{GHz}) carrier.
In contrast, there is a period from approximately
80 to 170 seconds where the SINR from both
mmWave cells is low.  
This time period corresponds to a 
segment of the route where
the UE is in \gls{nlos} to both
mmWave cells.

The resulting TCP end-to-end throughput is shown in Fig.~\ref{fig:herald}(b). We see that the LTE rate is more continuously available. However,
the maximum rate is limited to $\sim$\SI{36}{Mbps}
corresponding to the maximum modulation
and coding scheme (MCS) with a \SI{10}{MHz}
bandwidth.  In contrast, the mmWave system
can obtain the full \SI{120}{Mbps} rate,
but the rate falls below the LTE rate during some
parts of the NLOS segment.



\begin{figure*}
	\begin{subfigure}{0.5\linewidth}
		\centering
		\includegraphics[width=0.95\linewidth]{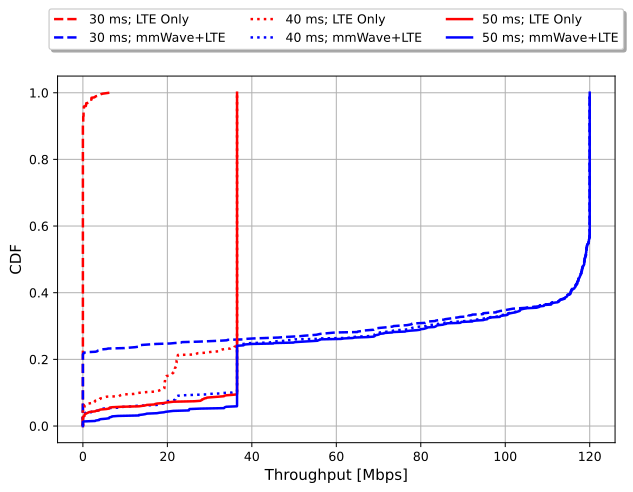}
		\caption{CDF of the delay-constrained end-to-end throughput. }
	\end{subfigure}
	\begin{subfigure}{0.5\linewidth}
		\centering
		\includegraphics[width=0.95\linewidth]{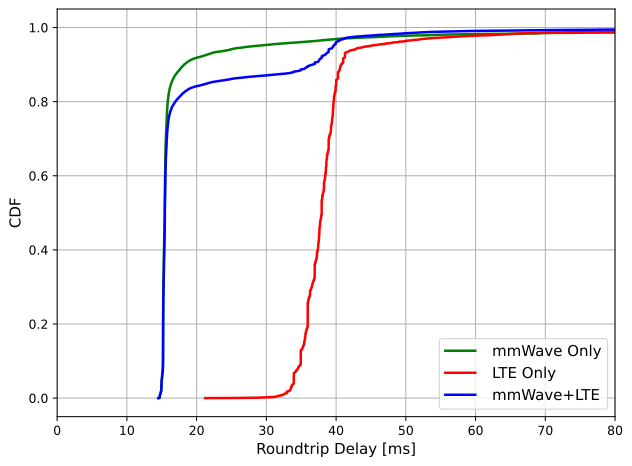}
		\caption{CDF of the unconstrained end-to-end roundtrip delay in milli-seconds.}
	\end{subfigure}
    \caption{Aggregated simulation results for a hypothetical user (\gls{ue}) walking across the four 
    sites in Manhattan. The delay-constrained
    throughputs in the left plot 
    are plotted for different rountrip delay constraint values $D_{\rm max} = 30, 40, 50\, \si{ms}$.}
    \label{fig:ns3_results}
\end{figure*}


The simulation results for each of the four 
scenarios have been aggregated and depicted in Fig.~\ref{fig:ns3_results}.  Fig.~\ref{fig:ns3_results}(a) plots what we will
call the \textit{delay-constrained throughput},
which is calculated as follows.  We divide
the time into intervals of $T$ seconds,
where $1/T = 30$\,\si{Hz} corresponds to the expected
video frame rate ($T$ is the frame interval).  For each TCP packet transmitted in the interval, we  measure its uplink
delay from the UE to the edge server application.
We also measure the downlink delay for each feedback packet transmitted
in that interval as well.  Note that these delays
contain all the air-link and core network delays.
For a given delay constraint, $D_{\rm max}$,
we define the delay-constrained throughput as $b/T$,
where $b$ is the number of uplink bits transmitted
in the interval $T$ for which the uplink +
downlink delay $\leq D_{\rm max}$. 

The delay-constrained rate is computed separately
for the LTE and mmWave systems.  Since mmWave systems
are always deployed with a sub-6 GHz fallback carrier,
we also estimate the rate of mmWave+LTE system
as the maximum of these two rates under the same delay constraint.
Fig.~\ref{fig:ns3_results}(a) plots the 
delay constrained rates for the LTE and mmWave+LTE
systems under delay constraints of $D_{\rm max}=$\,
30, 40 and 50\,\si{ms}.

We see in Fig.~\ref{fig:ns3_results}(a)
that the LTE system achieves a peak uplink 
rate of approximately \SI{36}{Mbps}, and
attains over \SI{20}{Mbps} more than 90\% of
the time. However, this rate is only achievable
with a delay constraint of $D_{\rm max}=50$\,\si{ms}.
At a tighter delay constraint of $D_{\rm max}=40$\,\si{ms}, the delay-constrained rate of approximately \SI{36}{Mbps} is supported less than 80\% of the time; at $D_{\rm max}=30$\,\si{ms},
there are virtually no LTE data within this delay
constraint.  In contrast, the mmWave 5G
system coupled with \gls{lte} can obtain the peak rate of \SI{120}{Mbps}
at least 40\% of the time, even under a delay
constraint of $D_{\rm max}=30$\,\si{ms}.
However, for approximately 25\% of the time,
the delay-constrained rate of the mmWave+LTE
system is similar to that of the LTE system
owing to the fact that the mmWave only coverage
is not always available and falls back to the LTE
carrier.

To understand the delay differences,
Fig.~\ref{fig:ns3_results}(b) plots the CDF
of the delays of packets without any delay constraint.  We see that for the
mmWave system, the minimum delay is $\approx$\,\SI{15}{ms} which includes two times
the core network delay of $D_{\rm core}=$\,\SI{5}{ms}
along with an addition \SI{5}{ms} for the transmission
of the uplink and downlink data.
The LTE packet delays are generally higher.
Although the core network delay is assumed to be
the same, the LTE frame structure as well as the lower throughput results
in a higher air-link delay.

Finally, it is useful
to compare these results with the 
URLLC requirements
of 5G. As mentioned in the Introduction, the URLLC design goal is to achieve
air-link latencies of 1 to \SI{10}{ms} \cite{li20185g, popovski2019wireless}.
When the UE has 5G mmWave connectivity, we see a median
delay of \SI{15}{ms}, which is consistent
with an air-link round-trip
latency of \SI{5}{ms} along with our
assumed core network delay of \SI{5}{ms} each way.
However, our study also includes blockage
and environments where the 5G coverage is not uniformly
available.  As a result, the UE must occasionally 
fall back
to  LTE links where the delay is higher
and the bandwidth cannot sustain the peak rates.
In these cases, the overall delay grows significantly,
beyond the 5G URLLC levels.

\newcommand{\badcell}[1]{\cellcolor{red!20}{#1}}
\newcommand{\medcell}[1]{\cellcolor{orange!20}{#1}}
\newcommand{\goodcell}[1]{\cellcolor{green!20}{#1}}

\begin{table*}[t]
\begin{center}
\begin{tabular}{|>{\raggedright}m{2.5cm}|>{\raggedright}m{1.3cm}|>{\raggedright}m{1.3cm}|
>{\raggedright}m{1.3cm}|>{\raggedright}m{1.3cm}|>{\raggedright}m{1.3cm}|>{\raggedright}m{1.3cm}|>{\raggedright}m{3.5cm}| }
\hline

 \tblhdr{Item} & 
\multicolumn{2}{c|}{\tblhdr{Local only}} &
 \tblhdr{LTE only} &
\multicolumn{2}{c|}{\tblhdr{mmWave+LTE}} & \tblhdr{Adaptive}
& \tblhdr{Remarks}\tabularnewline \hline

\multicolumn{8}{|l|}{\tblsubhdr{Video configuration and object detection performance} }
\tabularnewline \hline

Number of monocular cameras & \badcell{1} & \badcell{1} & \badcell{1} & \badcell{1} & \goodcell{4} & \goodcell{Variable (1-4)$^*$}  & 
$^*$1 camera if throughput $\leq$ \SI{26}{Mbps}, more cameras if throughput >26 Mbps
\tabularnewline \hline
Camera resolution & WVGA & 720P & 1080P & 
1080P & 1080P & mostly 1080P &
\tabularnewline \hline
wmAP (\%) for multiple objects  & \badcell{36.6} & \medcell{50.3}  & \goodcell{54.0} & 
\goodcell{54.0} & \goodcell{54.0} & \goodcell{54.0} avg$^\dagger$ &   See Table~\ref{tab:ComplexityVsResolution}
\tabularnewline \hline
AP (\%) for person  & 44.3 & 60.3 & 66.1 & 66.1 & 66.1 & 65.8 avg$^\dagger$ 
 & See Table~\ref{tab:ComplexityVsResolution}
\tabularnewline \hline
Detection range for person (meter) & 6 & 9 & 12 &12 &12 & mostly 12 & See Fig.~\ref{fig:standing_people_range}
\tabularnewline \hline


\multicolumn{8}{|l|}{\tblsubhdr{Bandwidth, Delay, and Availability} }
\tabularnewline \hline

Total uplink data rate (Mbps)  & 0 & 0  & 26 & 26 & 104 & variable & 
We assume each camera stream takes 26 Mbps except in the adaptive case. 
\tabularnewline \hline
Video frame delay (ms) & 33 & 33 & 33 & 33 & 33 & 33 & We assume video is captured at 30 Hz.
\tabularnewline \hline
Video encoding delay (ms)  & 0 & 0 & 17 & 17 & 17 & 17 & We assume that video encoding takes at most  1/60 (s) per frame, because today's smart phones can capture video at 60 Hz.  Local processing does not need to compress video.
\tabularnewline \hline
Inference time (ms) & 75 & 96 & 19 & 19 & 19 & variable & See Table~\ref{tab:ComplexityVsResolution}. We assume multiple GPUs process separate frames simultaneously when multiple camera data are uploaded.
\tabularnewline \hline
Median round-trip time (RTT) (ms) & - & - & 37 & 15 & 15 & 15 & 
See Fig.~\ref{fig:ns3_results}(b).
\tabularnewline \hline

Median total delay (ms) & \medcell{108} & \badcell{129} & \medcell{106} & \goodcell{84} & 
\goodcell{84}& \goodcell{84} & 
Frame delay+encoding delay+inference time+RTT
\tabularnewline \hline

\textcolor{black}{Availability with total delay $\leq 100$\,\si{ms} }& \badcell{0\%} & \badcell{0\%} & \badcell{0\%} & \medcell{75\%} &  \medcell{65\%} & \medcell{78\%} & \multirow{2}{3.5cm}{Computed from CDF of delay constrained throughput.  See text.}
\tabularnewline \cline{1-7}

\textcolor{black}{Availability with total delay $\leq 150$\,\si{ms}} & \goodcell{100\%} & \goodcell{100\%} & \goodcell{95\%} & \goodcell{97\%} & \medcell{67\%} & \goodcell{100\%} &
\tabularnewline \hline

\end{tabular}
\end{center}

\begin{tikzpicture}[xscale=1, yscale=1,every text node part/.style={align=left}]
    \newcommand{\textwid}{5.3cm}
    \footnotesize
    
\node [minimum width=1cm] (space) {};
\node[draw, right of=space, fill=red!20,
    minimum width=1.8cm, minimum height=0.5cm] (bad) {};    
\node[right of=bad,xshift=3cm,align=left,text width=6cm] {Performance value is poor};
\node[draw, below of=bad, yshift=0.4cm, fill=orange!20,
    minimum width=1.8cm, minimum height=0.5cm] (med) {};    
\node[right of=med,xshift=3cm,align=left,text width=6cm] 
    {Performance value is medium};
\node[draw, below of=med, yshift=0.4cm, fill=green!20,
    minimum width=1.8cm, minimum height=0.5cm] (good) {};    
\node[right of=good,xshift=3cm,align=left,text width=6cm] 
    {Performance value is good};    
    
\end{tikzpicture}

$^\dagger$The average wmAP and AP presented are for a total delay of $\leq 100$\,\si{ms}. These numbers are 53.9\% and 66.0\%, respectively for a total delay $\leq 150$\,\si{ms}.

There are many other feasible configurations, including  \\
(1) Processing 1080P video locally for increased detection accuracy, at a total delay of \SI{211}{ms}.\\
(2) Sending both views of each stereo camera or one view plus depth map, with increased uplink rate, and reduced availability.


\caption{Comparison summary of example configurations in different connectivity scenarios.} 

 
\label{tab:summary}

\end{table*}

\section{Performance Evaluation}
\label{sec:evaluation}

\subsection{Overview}
We now combine the video processing
requirements in Section~\ref{sec:video}
with the wireless simulation results in Section~\ref{sec:wireless} to assess the 
potential benefits of wireless offloading.
We consider three scenarios for edge connectivity: 
(1) \textbf{Local only},
where all the processing is performed on the
wearable; (2) \textbf{LTE only}, where edge processing
can be accessed by the LTE link; and (3)
\textbf{mmWave+LTE} where edge processing
can be accessed by LTE or mmWave, whichever
has the highest rate. 
For each such scenario, we consider different 
possible video options in terms of the
number of cameras, spatial resolution, and bit rate.
We can then use the wireless analysis 
in Section~\ref{sec:wireless} to 
assess the percent of time such video options
would be available based 
within a delay budget close to the
target of \SI{100}{ms}.  
Although there are a large number of possible video configurations, in the sequel,
we will focus on the options in Table~\ref{tab:summary} as these provide a good
demonstration of the capabilities of 
the system.  The table also highlights some of the key
values in red, orange, and green to draw attention
to the performance that are relative poor, medium,
or good.  
We also examine an adaptive offloading strategy, which switches between edge and local computing and furthermore adapts the video resolution based on the wireless link throughput when edge computing is chosen.  The remaining sub-sections will
describe the details of these options and their
analysis.

\subsection{Video Configurations}
In Table~\ref{tab:summary}, for edge 
computing with mmWave or LTE
connectivity, we have considered the case where the
video from each monocular camera would be delivered at 1080P spatial resolution, \SI{30}{Hz} temporal resolution, at \SI{26}{Mbps}. 
Based on the video analysis in Section~\ref{sec:video}, 
this configuration provides a high object detection 
accuracy and good
detection range -- see the wmAP and detection range
rows in Table~\ref{tab:summary}. Going beyond 1080P resolution and 26 Mbps brings only very slight gains, and yet processing 2.2K video will consume substantially more computation time. 

For the local processing scenario, 
we have considered only WVGA and 720P. 
With local processing, the inference time for higher
spatial resolution 1080P (see Table~\ref{tab:ComplexityVsResolution}) would substantially exceed the delay budget.
 As shown in Table~\ref{tab:summary},
this lower resolution results in both a lower
object detection accuracy and reduced object
detection range.

\subsection{Delay Analysis}
The delay computations in Table~\ref{tab:summary}
consider four components:
\begin{itemize}
    \item \textit{Video frame delay} which is the interval of one video frame (i.e., the inverse
    of the frame rate).  For edge 
    computing, the video frame interval needs to be considered because an object may appear any time between two adjacent frames.
    
    \item \textit{Video encoding delay} which is the time to encode the video for edge computing.
    
    \item \textit{Round-trip time (RTT)}
    which is the total time to transmit the packets
    from the wearable to the edge server and back.
    
    \item \textit{Inference time} which is the 
    time for the object detection network (either
    local or edge) to compute the detection results.
\end{itemize}
In our analysis, the video frame delay, encoding delay,
and inference time are fixed.  The only variable
component is the RTT. Table~\ref{tab:summary}
shows the median RTT and the corresponding median
total time.
We see that mmWave+LTE offers dramatically lower median
RTT of \SI{15}{ms} relative to the median RTT in LTE
only of \SI{37}{ms}.  Recall that the RTT includes
$2D_{\rm core}=10$\,\si{ms} of assumed delay 
from the base station (gNB or eNB) through the core
network to the edge server and back.
Note that with direct local processing, only the
frame delay and inference time are required since there is no video encoding or communication.


As suggested in \cite{liu2019edge}, to avoid  detection mismatch, we will assume that the local processor runs a simple object tracking algorithm that predicts the locations of  objects detected for the  last frame for which the edge server detection results were fed back. For example, if the total delay for edge processing is twice of the frame interval, at the time when frame $t$ is captured, detection results for frame $t-2$ will be used as the reference, the motion between frame $t$ and frame $t-2$ will be used to predict the locations of these detected objects in frame $t$. In the mean time, frame $t$ will be delivered to the edge server with its processing results to be fed back at time when frame $t+2$ will be captured. The motion vectors between frames generated for video compression can be leveraged for local object tracking. Such an approach should be able to track small movements of previously detected objects within a few frames, while also reporting any newly appearing objects within the 100 ms delay.


\subsection{Supportable Video Streams Under Different Delay Constraints}

\begin{figure*}
	\begin{subfigure}{0.5\linewidth}
		\centering
		\includegraphics[width=0.95\linewidth]{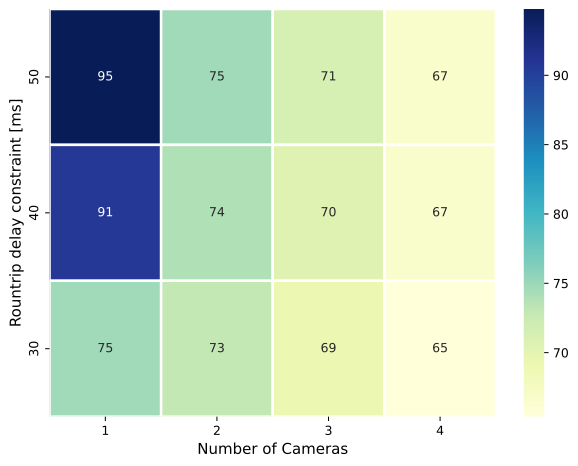}
		\caption{\Gls{mmw} + \gls{lte} case, all cameras with 1080P resolution @ 26 Mbps.}
	\end{subfigure}
	\begin{subfigure}{0.5\linewidth}
		\centering
		\includegraphics[width=0.95\linewidth]{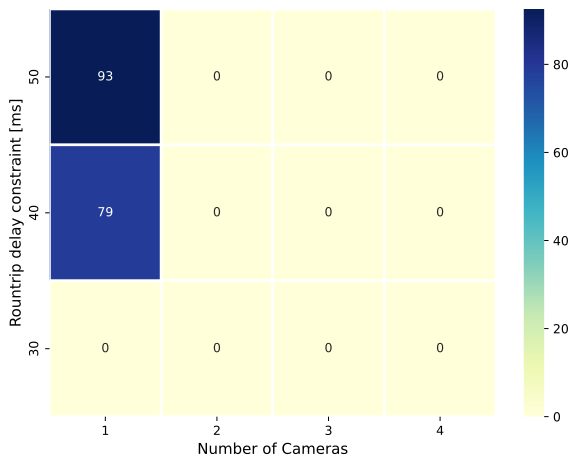}
		\caption{\Gls{lte} only case, all cameras with 1080P resolution @ 26 Mbps.}
	\end{subfigure}
	\quad
	\begin{subfigure}{\linewidth}
		\centering
		\includegraphics[width=0.475\linewidth]{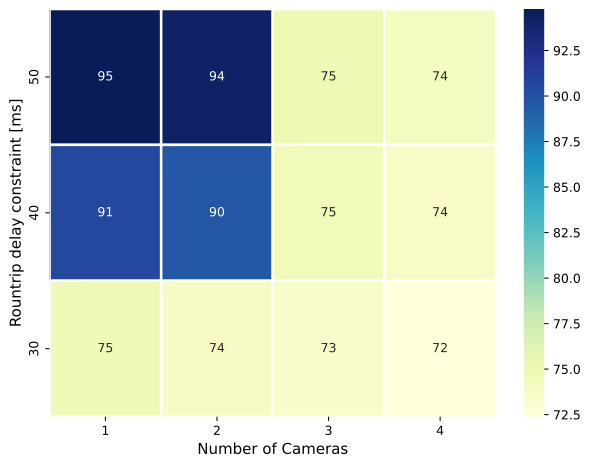}
		\caption{\Gls{mmw} + \gls{lte} case, camera $1$ @ 26 Mbps, cameras 2, 3, 4 @ 10 Mbps.}
	\end{subfigure}
    \caption{Heat-map of probabilities of supporting different numbers of cameras under  different roundtrip delay constraints.}
    \label{fig:heatmap}
\end{figure*}

From the wireless simulation results in Figure~\ref{fig:ns3_results}, 
we derive the probability of supporting one or more camera streams under different round-trip delay constraints.
These are plotted as heat maps in Fig.~\ref{fig:heatmap}.
Fig.~\ref{fig:heatmap}(a) shows that, with multi-connectivity using both \gls{mmw} and \gls{lte} links, we can support one video stream over 75\% of the time, and support two streams over 73\% of the time, under the the roundtrip delay constraint of \SI{30}{ms}.
All four cameras can be supported 65\% of the time. Furthermore, we can support one and four  cameras with high availability (91\% and 67\%, respectively) if the delay constraint is relaxed to \SI{40}{ms}.
On the other hand, with the \gls{lte} link only (Figure.~\ref{fig:heatmap}(b)) and a round-trip
delay of \SI{40}{ms}, 
the availability for supporting one and four cameras drops to 79\% and 0\%, respectively, since the peak \gls{lte} rate is approximately \SI{36}{Mbps}. 
The \gls{lte} link can sustain one camera with high probability (93\%) only if the delay constraint is relaxed to \SI{50}{ms}.

For any configuration,
we can also compute the probability that the total 
delay will meet a certain delay target.  For example,
from Section~\ref{sec:delay_requirements}, the
estimated total delay requirement is \SI{100}{ms}.
This is not met by local processing.  When using offloading,
the video frame delay, encoding delay, and inference
take a total of $33+17+19=69$\,\si{ms}, so there
would be $D_{\rm max}=100-69=31$\,\si{ms} for the RTT.
Similarly,
if we relax the total delay requirement to 
\SI{150}{ms}, the communication delay constraint
would be $D_{\rm max}=150-69=81$\,\si{ms}.
The availability numbers listed in the final two 
rows of Table~\ref{tab:summary} are  the percentage of
time the delay-constrained throughput meets the minimum
uplink data rate requirement at the RTT of 31 and 81 \si{ms}, respectively. 
The availability percentages for these two delay constraints are not shown in Fig.~\ref{fig:heatmap}, but we have extracted the numbers from the wireless simulation results in a similar manner.
Note that the availability for RTT of \SI{31}{ms} is the practically the same as for \SI{30}{ms}.  Using Fig.~\ref{fig:heatmap}, one can also find the corresponding availability for total delays of \SI{110}{ms} (RTT of \si{40}{ms}) or \si{120}{ms} (RTT of \si{50}{ms}).

In practice, given the limited total throughput, it may be better to only upload the front facing camera stream at a high rate (for the highest detection accuracy), and use a lower rate for other cameras (side and back facing) to enhance situational awareness.
As an example, Fig.~\ref{fig:heatmap}(c) shows the probability of supporting one camera at the full rate
of \SI{26}{Mbps} and additional cameras at \SI{10}{Mbps} each.   In this case, with mmWave+LTE
connectivity and a total delay 100 \si{ms}, we could increase the probability to support  4 cameras from 65\% to 72\% .




\subsection{Adaptive Offloading}

Given the variability in availability,
particularly in mmWave, it is natural 
to consider a strategy
that adaptively selects the rate,  
number of cameras,
and whether to use local or remote
processing based on the available 
uplink bandwidth.
For example, as a simple
strategy, when the available throughput is $\geq$\SI{26}{Mbps}, we can transmit
one or more cameras.
When the throughput is lower than \SI{26}{Mbps}, we can still offload the video for edge computing, but at lower
rates, enabling adaptive switching between camera number and video quality based on bandwidth constraints and functional need (refer to 
Fig.~\ref{fig:heatmap} as an example).
From Fig.~\ref{fig:wmAPvsRate}, when the throughput is between \SI{6}{Mbps} and \SI{26}{Mbps}, the system should  deliver the video  at 1080P but at a lower rate, leading to proportionally lower detection accuracy. When the rate is between \SI{1}{Mbps} and \SI{6}{Mbps}, the system should  upload the video at 720P resolution. 
In the ``adaptive'' column of Table~\ref{tab:summary}, the detection accuracy is derived by assuming an average detection accuracy of 51.5 and 63.2 for multi-object and person, respectively, when the bit rate is between 6 and 26 Mbps (which occurs 2\% of the time with a delay constraint of \SI{30}{ms}, from Fig.~\ref{fig:ns3_results}(a)); and an accuracy of 41.1 and 49.7, respectively, when the bit rate is between 1 Mbps and 6 Mbps (which occurs 1\% of the time).
The overall availability for adaptive offloading under total \SI{100} ms delay is the probability that the throughput is $\geq 1$ {Mbps} at a RTT constraint of \SI{30}{ms}. 

Under the relaxed total delay constraint of \SI{150} ms, when the throughput is between 10 and 26 Mbps (with probability of 2.2\%), the system should still upload the 1080P video. When the throughput is below \SI{10} Mbps (with probability of 0.8\%), the wearable could locally process the uncompressed video at  720P video resolution for better detection performance (see Fig.~\ref{fig:wmAPvsRate}). Therefore, the availability is 100\%. The average accuracy would be 53.92\% and 66.00\%, for multi-object and person, respectively.



\section{Conclusions and Future Work} \label{sec:conclusion}

Mobile edge computing coupled with the high data
rate capabilities of mmWave holds significant promise for accessing powerful video analytics by wearable devices. 
We have assessed the
feasibility of such capabilities for a 
advanced smart wearable with multiple high-resolution cameras where the wireless and video requirements
are particularly demanding.  
Several new elements were required in the analysis
including  developing a large labeled video data set,
evaluation of object detection algorithms at variable resolutions and bit rates, and detailed and high accuracy wireless simulations with ray tracing.  
Overall, wireless simulations provide a high level
of realism and can 
identify the key limitations in 
high data rate edge computing.
These tools can be applied in other applications
and may prove valuable 
as video processing and 
spatial intelligence becomes more widely-used
in mobile scenarios.

For the \VISION application, our simulation
results suggest that at bandwidths and loading 
similar to current deployments, systems in
traditional sub-6-GHz bands combined with low delay 
mobile edge computing can provide gains by
offloading camera data
a large fraction of time, improving the accuracy 
and detection range.  However, meeting the
end-to-end delay requirements of \SI{100}{ms} is challenging.  
The mmWave bands can reduce the delay to meet these requirements 
and provide additional 
capabilities including multiple
cameras at high resolution.
However, due to blockage and the
limited range
of mmWave signals, the peak performance is not uniformly available at typical 
cell-site densities.
Thus, fall back to lower frequency carriers and local
processing combined with adaptation in the video resolution
and number of camera streams  will be required.

In the current work, we have abstracted this adaptation
by assuming that the number of cameras and their bit rate can be adjusted to the available throughput. 
An obvious line of future work is to actually simulate
a particular adaptive video application over wireless links and assess
its performance.  Additionally, we have simply relied on the pre-trained YOLO network that
was trained on uncompressed low-resolution images.
A second line of work is to train multiple
detection networks for different resolutions and bit rates or a single network that can perform well across resolutions and bit rates. 
More ambitiously, 
one can also consider new compression
schemes that are trained end-to-end with
object detection accuracy as the goal, as opposed to the standard compression algorithms, which are optimized for image reconstruction.

\paragraph*{Acknowledgements}
This work was supported in part by the NSF grant 1952180 under the Smart and Connected Community program
as well as the industrial affiliates of
NYU WIRELESS.
In addition,
Azzino, Mezzavilla and Rangan were supported by
NSF grants 1925079, 1564142, and 1547332 
and the Semiconductor Research Corporation (SRC). 
Yuan and Wang were also supported by NSF grant 2003182. Yu Hao was also partially supported by NYUAD Institute (Research Enhancement Fund - RE132).
Special thanks are also given to Remcom that provided the Wireless Insite  software and GeoPipe that
provided the 3D models.


\bibliographystyle{IEEEtran}
\bibliography{bibl}

\begin{IEEEbiography}[{\includegraphics[width=1in,height=1.25in,clip,keepaspectratio]{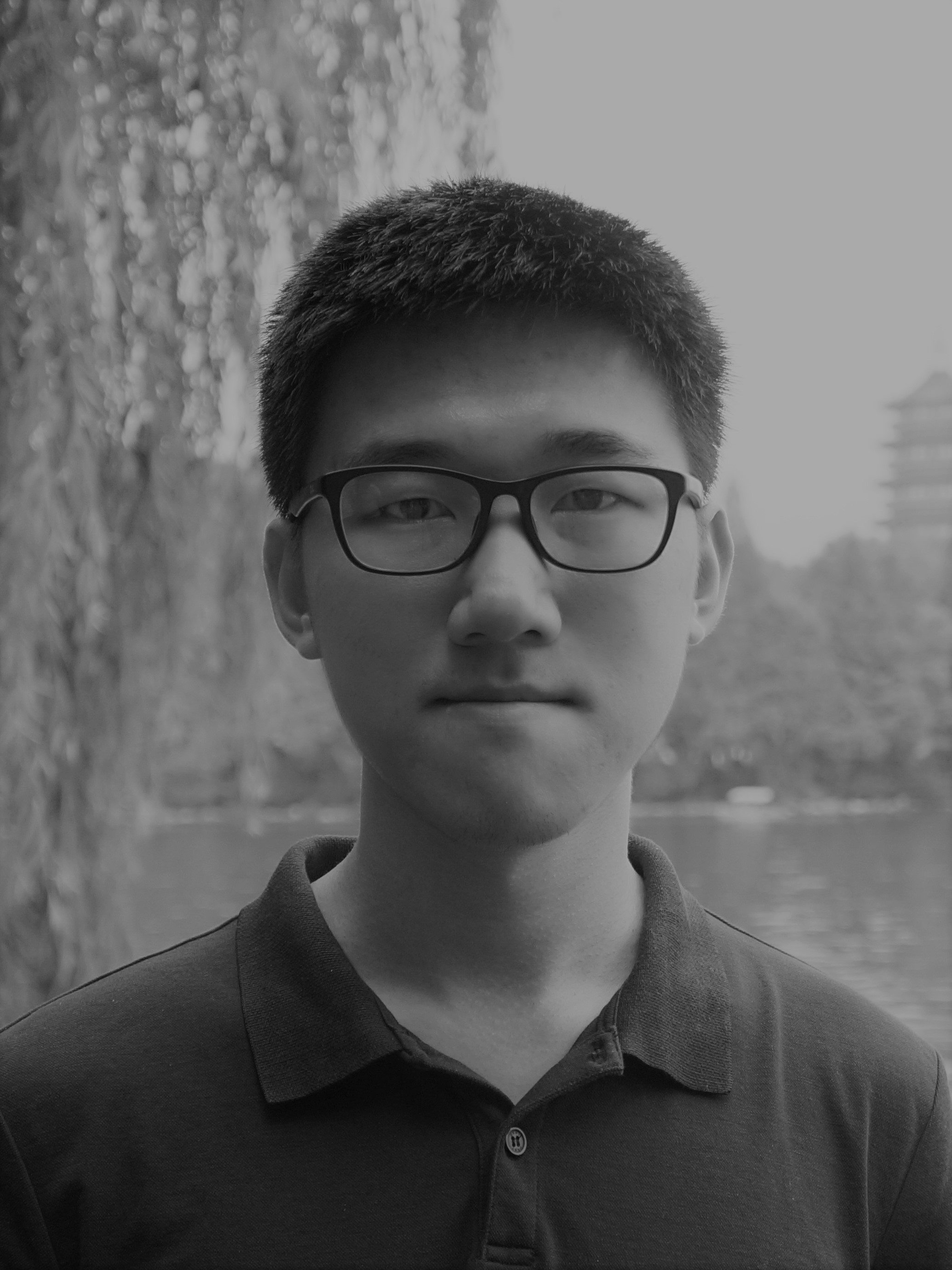}}]{Zhongzheng (Jacky) Yuan} received the B.S. and M.S. in electrical engineering from the New York University Tandon School of Engineering, Brooklyn, NY, USA, in 2019 and 2020, respectively. He had held internship positions in Media Lab at Tencent America, and Incubation Lab at Dolby Laboratories.

He is currently a Ph.D. candidate in electrical engineering at New York University, under the supervision of Prof. Yao Wang. His research interests include image/video compression, computer vision, and deep learning.
\end{IEEEbiography}

\begin{IEEEbiography}[{\includegraphics[width=1in,height=1.25in,clip,keepaspectratio]{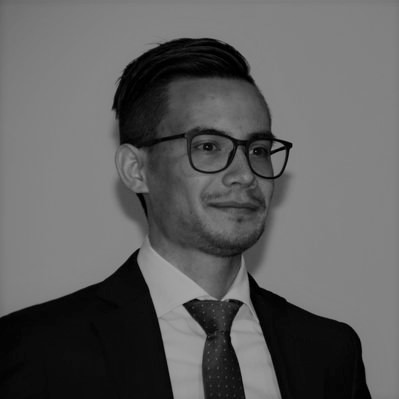}}]{Tommy Azzino}  received the B.Sc. and M.Sc. in information engineering and telecommunications engineering, respectively, from the University of Padova, Italy, in 2016 and 2019. He is currently pursuing the Ph.D. degree in electrical and computer engineering with New York University Tandon School of Engineering, Brooklyn, NY, USA, under the supervision of Prof. S. Rangan. He held visiting research positions with the National Institute of Standards and Technology (NIST), Department of Commerce, Gaithersburg, MD, USA, in 2018, Nokia Bell Labs France, Paris, France, in 2019.

He took part to the 2018 edition of the Seeds for the Future Project promoted by Huawei Technologies in Shenzhen, PRC, in 2018. His research interests include 5G mobile cloud/edge computing, 5G network simulation and prototyping, MAC and network-layer resource allocation, wireless communications, and machine learning  for wireless communications.
\end{IEEEbiography}

\begin{IEEEbiography}[{\includegraphics[width=1in,height=1.25in,clip,keepaspectratio]{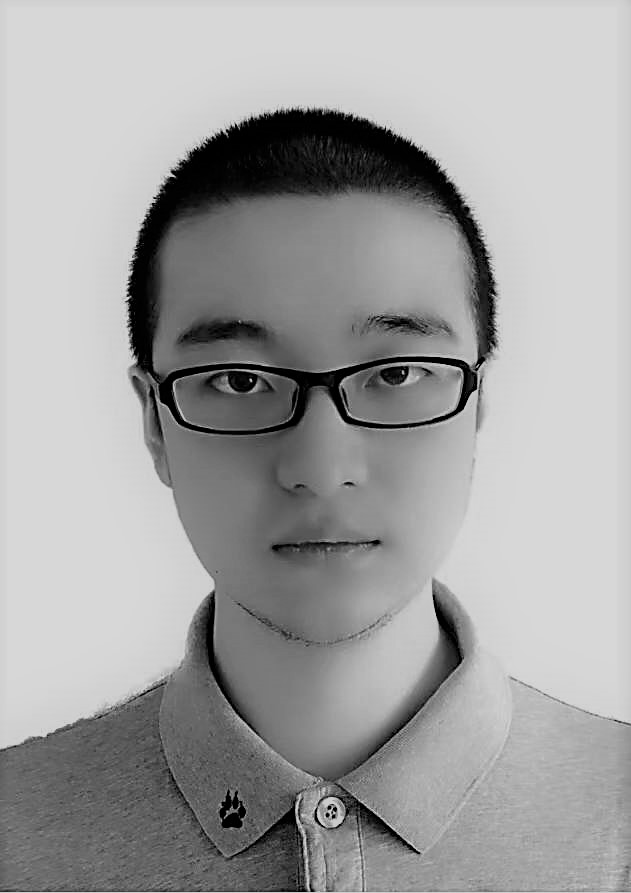}}]{Yu Hao} received his B.Eng degree in Computer Science and Technology from Beijing University of Technology, Beijing, China in 2019. He received an M.S degree in Computer Science from New York University, New York, USA in 2021. He is currently pursuing the Ph.D. degree in Computer Science at New York University. His research interests include Computer Vision, Computer Graphics, and Deep Learning. 
\end{IEEEbiography}

\begin{IEEEbiography}[{\includegraphics[width=1in,height=1.25in,clip,keepaspectratio]{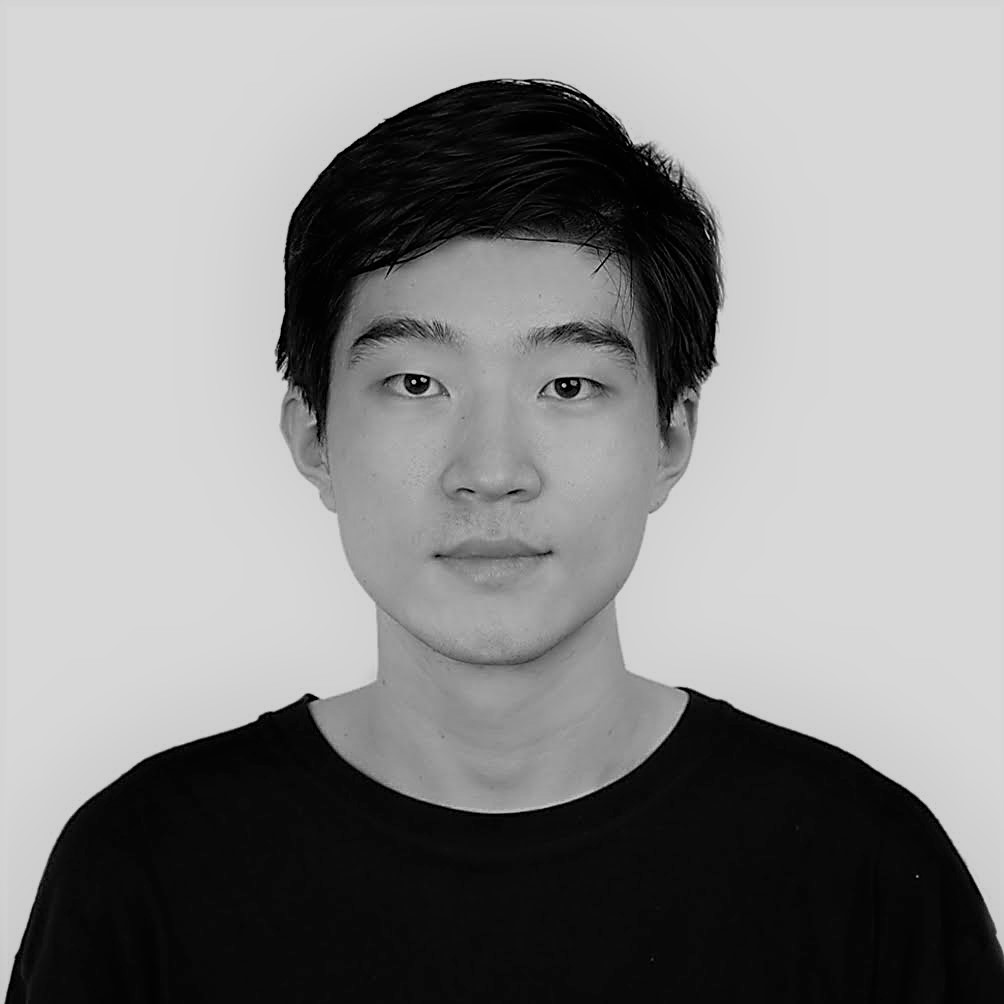}}]{Yixuan Lyu} was born In Beijing, China in 1996. He received a B.E. degree in automation from Nanjing University of Science and Technology in 2019 and now pursuing a master’s degree in New York University majoring in Computer Engineering. His research interests include video/image compression methods, computer vision, and deep neural networks.
\end{IEEEbiography}

\begin{IEEEbiography}[{\includegraphics[width=1in,height=1.25in,clip,keepaspectratio]{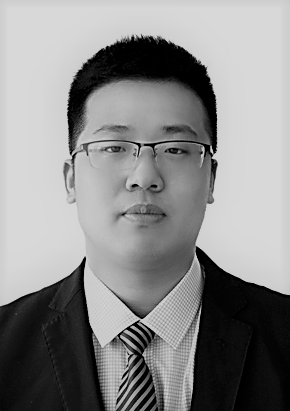}}]{Haoyang Pei} was born In Yuncheng, Shanxi, China in 1995. He received a B.E. degree in electrical engineering and its automation from North China Electric Power University in 2017 and now pursuing a master's degree in New York University majoring in Computer Engineering. His research interests include video/image compression, computer vision, and deep learning.
\end{IEEEbiography}

\begin{IEEEbiography}[{\includegraphics[width=1in,height=1.25in,clip,keepaspectratio]{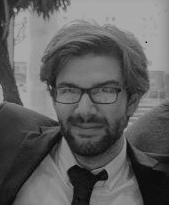}}]{Alain Boldini} received B.Sc. degree in Aerospace Engineering and M.Sc. degree in Aeronautical Engineering from Politecnico di Milano, Milan, Italy, in 2015 and 2017, respectively, and M.Sc. in Aerospace Engineering from Politecnico di Torino, Turin, Italy, in 2018.

He is currently a PhD candidate in Mechanical Engineering at the New York University Tandon School of Engineering. He is involved in research in smart advanced materials, assistive technologies, and dynamical systems.
\end{IEEEbiography}

\begin{IEEEbiography}[{\includegraphics[width=1in,height=1.25in,clip,keepaspectratio]{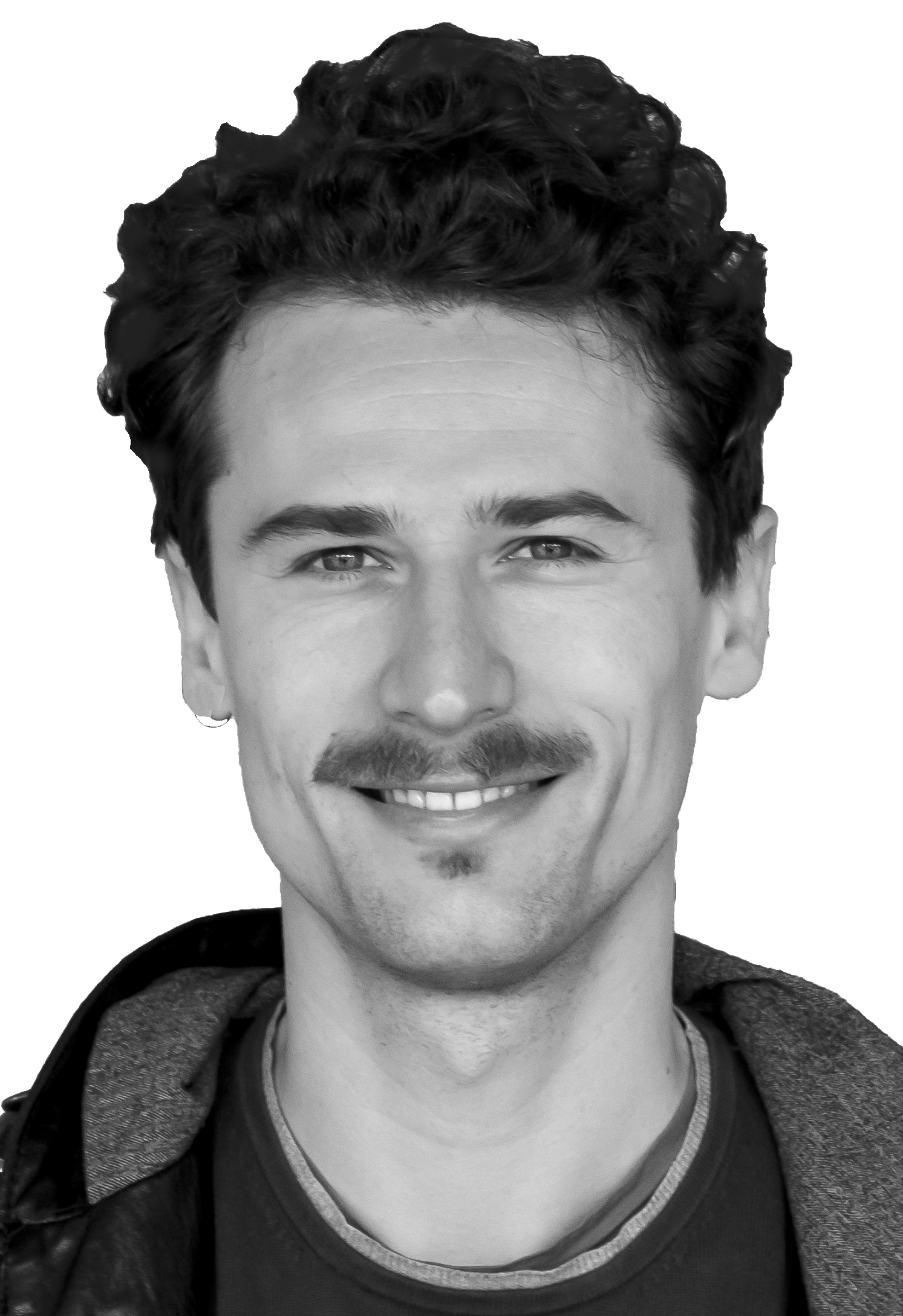}}]{Marco Mezzavilla} received the B.A.Sc., the M.Sc., and the Ph.D. at the University of Padua, Italy, all in Electrical Engineering. He held visiting research positions at the NEC Network Laboratories in Heidelberg (Germany, 2009), at the Telematics Department at Polytechnic University of Catalonia (UPC) in Barcelona (Spain, 2010), and at Qualcomm Research in San Diego (USA, 2012).

He joined New York University (NYU) Tandon School of Engineering in 2014, where he is currently a Research Faculty and leads multiple mmWave-related research projects, mainly focusing on 5G and beyond PHY/MAC design. He is serving as reviewer for many IEEE and ACM conferences, journals, and magazines. His research interests include design and validation of communication protocols and applications for 5G and beyond broadband wireless technologies, millimeter-wave communications, multimedia traffic optimization, radio resource management, spectrum sharing, convex optimization, cognitive networks, and novel network paradigms. He is a Senior Member of the IEEE. 
\end{IEEEbiography}

\begin{IEEEbiography}[{\includegraphics[width=1in,height=1.25in,clip,keepaspectratio]{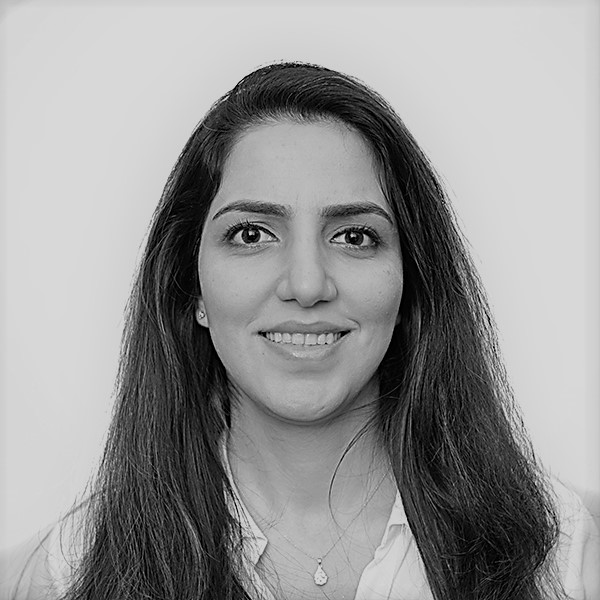}}]{Mahya Beheshti} received her M.D. degree from Gulf Medical University, School of Medicine, Ajman, UAE.

She is a Physician-scientist at NYU Langone Medical Center’s Rusk Rehabilitation. She is also pursuing the Ph.D. degree in Mechanical and Aerospace Engineering Department, NYU Tandon School of Engineering. She is the author of 3 book chapters, more than 17 peer-reviewed articles and 13 poster presentations. She has also given multiple talks. Her research interests lie within the realm of neurorehabilitation, human-machine interface, specifically EEG-based pattern recognition and wearable technologies.

Dr. Beheshti was a recipient of Distinguished Alumni Award of Healthcare Academia from Gulf Medical University and was chosen as Intern of the Year during her medical training. She is a member of American Society of Neuroradiology (ASNR), American Medical Association (AMA), National Multiple Sclerosis Society, and Association of Academic Physiatry (AAP). She is IEEE student member (IEEE RAS, IEEE EMBS, and IEEE WIE).
\end{IEEEbiography}

\begin{IEEEbiography}[{\includegraphics[width=1in,height=1.25in,clip,keepaspectratio]{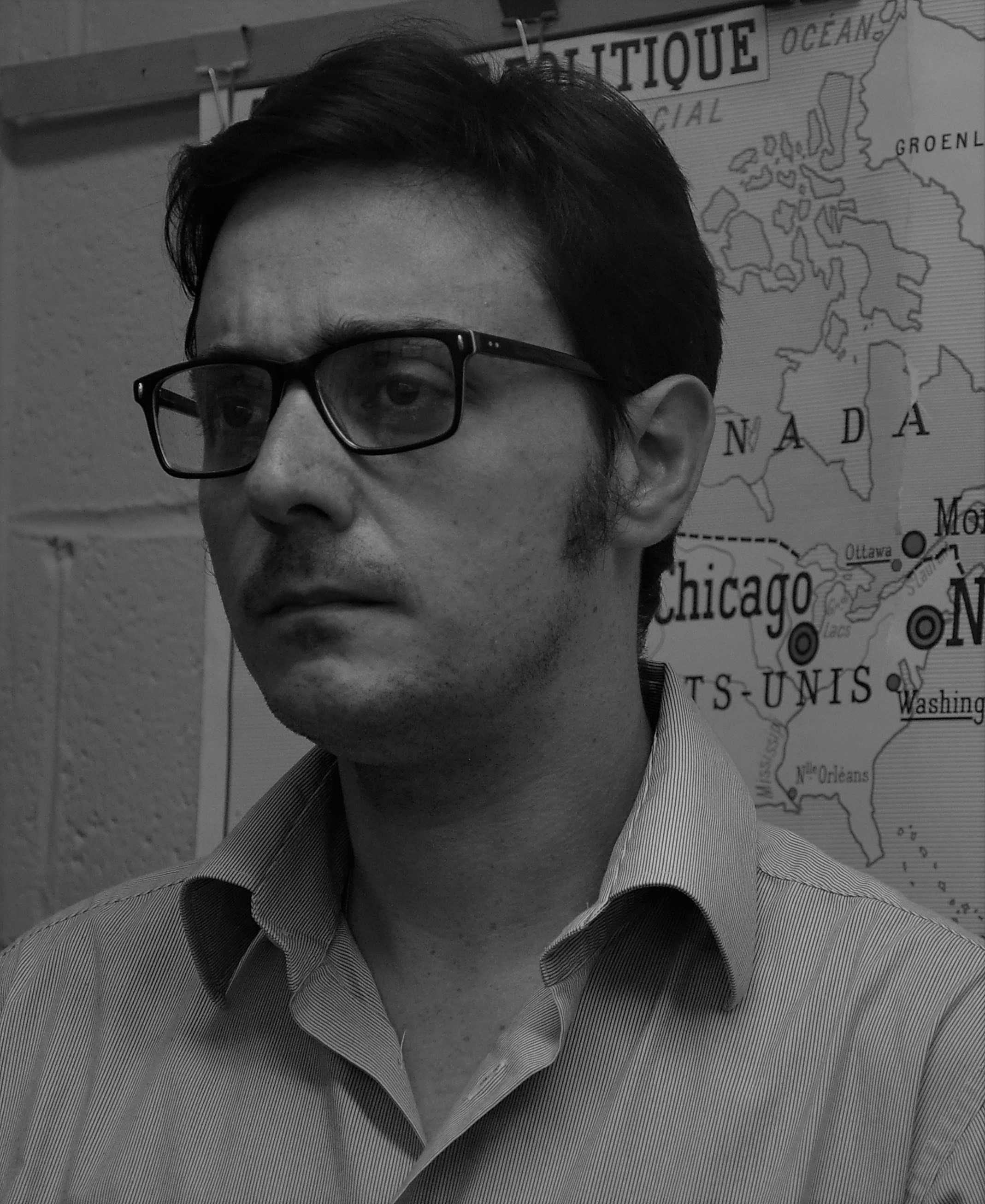}}]{Maurizio Porfiri} (Fellow, IEEE) received the M.Sc. and Ph.D. degrees in engineering mechanics from Virginia Tech, Blacksburg, VA, USA, in 2000 and 2006, respectively, and the “Laurea” (Hons.) degree in electrical engineering and the Ph.D. degree in theoretical and applied mechanics from the Sapienza University of Rome, Rome, Italy, and the University of Toulon, Toulon, France (under a dual-degree program between Italy and France), in 2001 and 2005, respectively.

He is currently an Institute Professor with the New York University Tandon School of Engineering, Brooklyn, NY, USA, having appointments with the Center for Urban Science and Progress, Department of Mechanical and Aerospace Engineering, Department of Biomedical Engineering, and Department of Civil and Urban Engineering. He is the Director of the Dynamical Systems Laboratory. He is involved in conducting and supervising research on dynamical systems theory and applications. He has authored more than 350 journal publications. He has been included in the “Brilliant 10” list of Popular Science in 2010. His research featured in major media outlets, including CNN, NPR, Scientific American, and Discovery Channel.

Dr. Porfiri is the recipient of the National Science Foundation CAREER Award, the Outstanding Young Alumnus Award by the College of Engineering, Virginia Tech, the American Society of Mechanical Engineers (ASME) Gary Anderson Early Achievement Award, the ASME DSCD Young Investigator Award, and the ASME C.D. Mote, Jr. Early Career Award. His other significant recognitions include invitations to the Frontiers of Engineering Symposium and the Japan–America Frontiers of Engineering Symposium organized by the National Academy of Engineering He has served on the Editorial Board of the ASME Journal of Dynamics Systems, Measurements and Control, the ASME Journal of Vibrations and Acoustics, Flow, the IEEE CONTROL SYSTEMS LETTERS, the IEEE TRANSACTIONS ON CIRCUITS AND SYSTEMS I: REGULAR PAPERS, and Mechatronics. He is a Fellow of the ASME.
\end{IEEEbiography}

\begin{IEEEbiography}[{\includegraphics[width=1in,height=1.25in,clip,keepaspectratio]{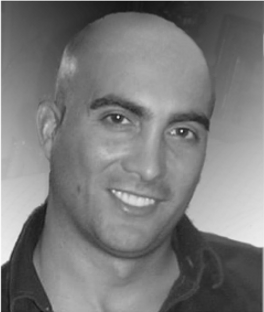}}]{Todd E. Hudson} is an Assistant Professor of Rehabilitation Medicine at New York University’s Grossman School of Medicine, holding cross-appointments in Neurology, and also in the Department of Biomedical Engineering at the New York University Tandon School of Engineering.

Prof. Hudson is a Computational Neuroscientist whose research focuses on modeling sensory and motor systems, particularly with regard to eye and arm movements in healthy and disease states, movement planning, and spatial orientation. He received his PhD from Columbia University, and has authored over 50 peer reviewed articles, as well as the textbook ‘Bayesian Data Analysis for the Behavioral and Neural Sciences’ from Cambridge University Press.
\end{IEEEbiography}

\begin{IEEEbiography}[{\includegraphics[width=1in,height=1.25in,clip,keepaspectratio]{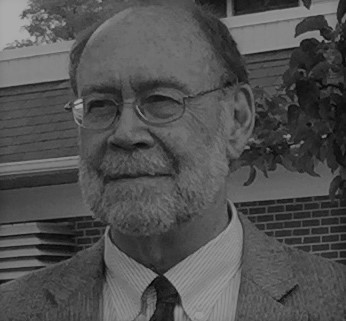}}]{William Seiple} received the B.S. and M.S. degrees in psychology from the Albright College, Reading, PA and the University of North Carolina, Greensboro, NC and the Ph.D. degree in zoology from the University of Illinois, Urbana, IL.

He is the Chief Research Officer at Lighthouse Guild, Research Professor of Ophthalmology at New York University School of Medicine,  and Adjunct Faculty at the Institut de la Vision, Paris. He is the author of two books chapters, and more than 150 peer-reviewed articles.

His research interests include development and assessment of functional interventions for people with vision loss, rehabilitation training, and visual electrophysiology and psychophysics of vision.
\end{IEEEbiography}

\begin{IEEEbiography}[{\includegraphics[width=1in,height=1.25in,clip,keepaspectratio]{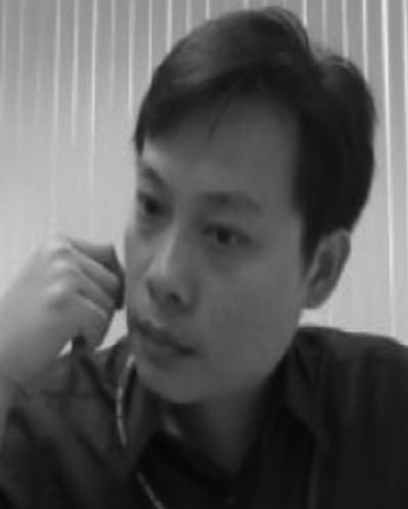}}]{Yi Fang} received the BS and MS degrees in biomedical engineering from Xi’an Jiaotong University, Xi’an, China, in 2003 and 2006, respectively, and the PhD degree in mechanical engineering from Purdue University, West Lafayette, Indiana, in 2011.

He is currently an assistant professor with the Department of Electrical and Computer Engineering, New York University Abu Dhabi, Abu Dhabi, United Arab Emirates. His research interests include three-dimensional computer vision and pattern recognition, large-scale visual computing, deep visual computing, deep cross-domain and cross modality multimedia analysis, and computational structural biology.
\end{IEEEbiography}

\begin{IEEEbiography}[{\includegraphics[width=1in,height=1.25in,clip,keepaspectratio]{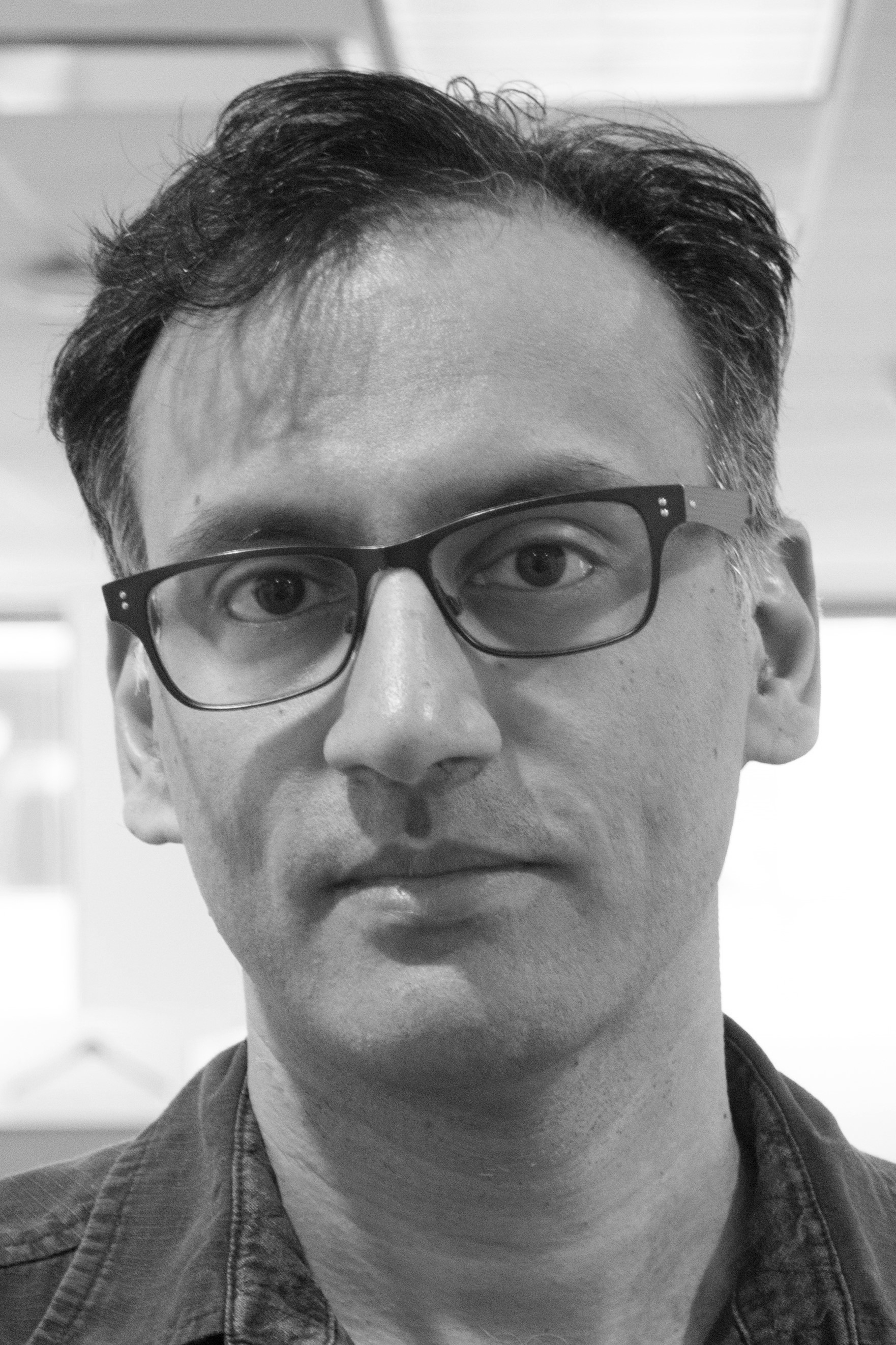}}]{Sundeep Rangan}  received the B.A.Sc. at the University of Waterloo, Canada and the M.Sc. and Ph.D. at the University of California, Berkeley, all in Electrical Engineering. He has held postdoctoral appointments at the University of Michigan, Ann Arbor and Bell Labs.  In 2000, he co-founded (with four others) Flarion Technologies, a spin-off of Bell Labs, that developed Flash OFDM, the first cellular OFDM data system and pre-cursor to 4G cellular systems including LTE and WiMAX. In 2006, Flarion was acquired by Qualcomm Technologies.  Dr. Rangan was a Director of Engineering at Qualcomm involved in OFDM infrastructure products. He joined NYU Tandon (formerly NYU Polytechnic) in 2010 where he is currently a Professor of Electrical and Computer Engineering.  He is a Fellow of the IEEE and the Associate Director of NYU WIRELESS, an industry-academic research center on next-generation wireless systems.
\end{IEEEbiography}

\begin{IEEEbiography}[{\includegraphics[width=1in,height=1.25in,clip,keepaspectratio]{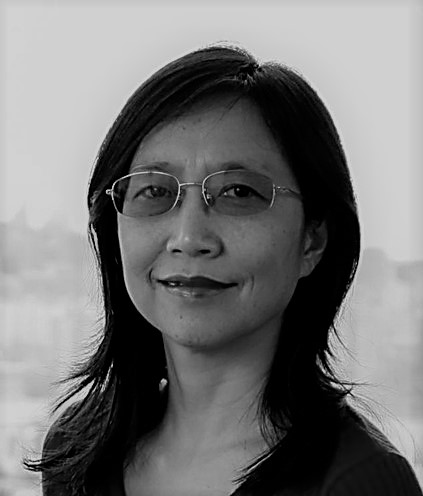}}]{Yao Wang} is a Professor at New York University Tandon School of Engineering (formerly Polytechnic University, Brooklyn, NY), with joint appointment in Departments of Electrical and Computer Engineering and Biomedical Engineering. She is also Associate Dean for Faculty Affairs for NYU Tandon since June 2019. Her research areas include video coding and streaming, multimedia signal processing, computer vision, and medical imaging. 

She is the leading author of a textbook titled Video Processing and Communications, and has published over 250 papers in journals and conference proceedings. She received New York City Mayor's Award for Excellence in Science and Technology in the Young Investigator Category in year 2000. She was elected Fellow of the IEEE in 2004 for contributions to video processing and communications. She received the IEEE Communications Society Leonard G. Abraham Prize Paper Award in the Field of Communications Systems in 2004, and the IEEE Communications Society Multimedia Communication Technical Committee Best Paper Award in 2011. She was a keynote speaker at the 2010 International Packet Video Workshop, INFOCOM Workshop on Contemporary Video in 2014,  the 2018 Picture Coding Symposium, and the 2020 ACM Multimedia Systems Conference (MMSys’20). She received the NYU Tandon Distinguished Teacher Award in 2016.
\end{IEEEbiography}

\begin{IEEEbiography}[{\includegraphics[width=1in,height=1.25in,clip,keepaspectratio]{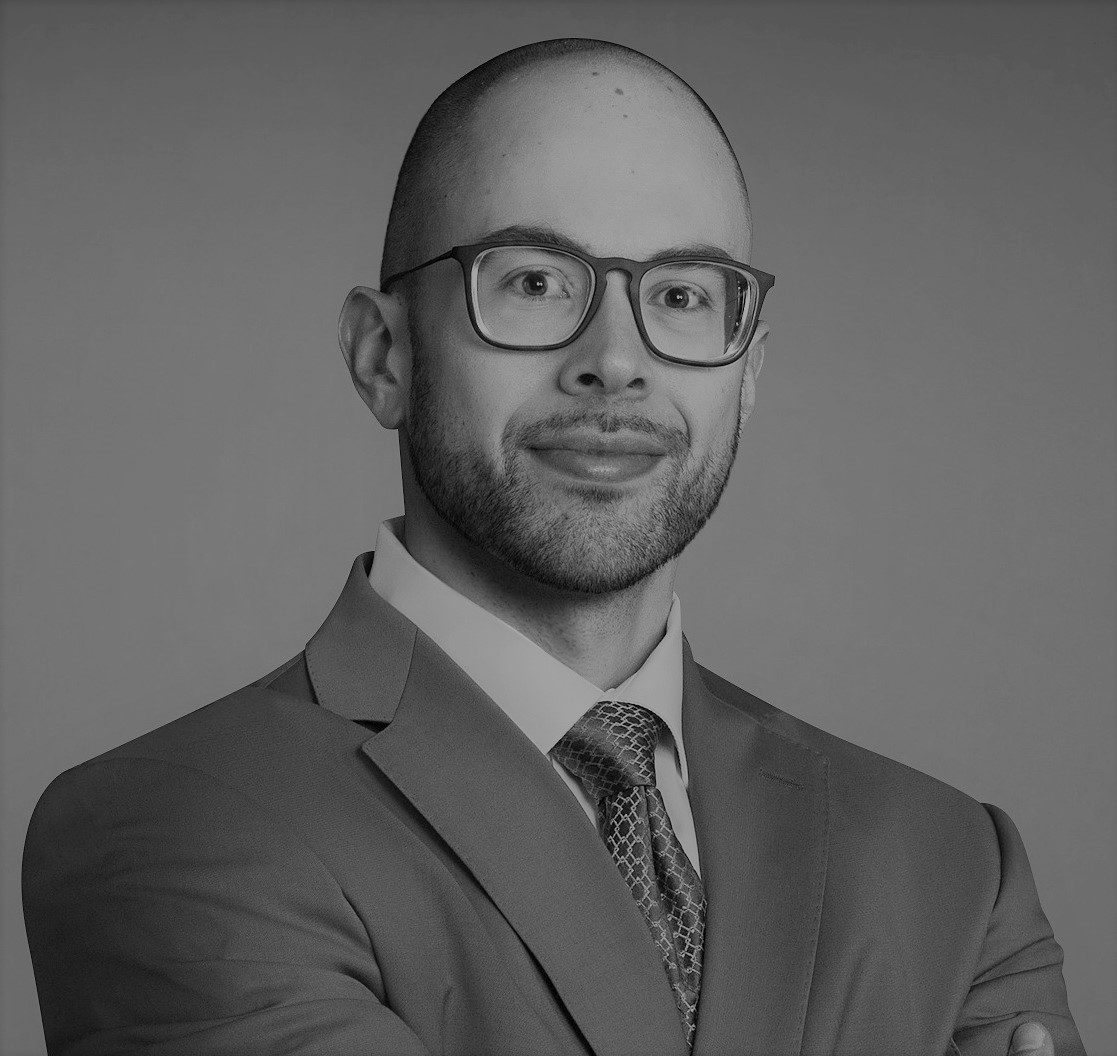}}]{John-Ross (JR) Rizzo} was born in Wayne, NJ, USA. He received his medical degree from New York Medical College, Alpha Omega Alpha Honors, on academic scholarship.

He is currently a physician-scientist at NYU Langone Medical Center’s Rusk Rehabilitation, where he serves as vice chair of Innovation and Equity for Physical Medicine and Rehabilitation with cross-appointments in the Department of Neurology and the Departments of Biomedical \& Mechanical and Aerospace Engineering at NYU-Tandon School of Engineering. He is also the Associate Director of Healthcare for the NYU Wireless Laboratory in the Department of Electrical and Computer Engineering at NYU-Tandon. He leads the Visuomotor Integration Laboratory (VMIL) and the REACTIV Laboratory (Rehabilitation Engineering Alliance and Center Transforming Low Vision), where his team focuses on assistive technology for the visually impaired and benefits from his own personal experiences with vision loss.  He is the author of 10 book chapters, more than 80 peer-reviewed articles and many poster presentations. His research interests lie within the realm of neurorehabilitation, assistive technologies, and health equity. 

Dr. Rizzo was awarded the prestigious Crain’s 40 under 40 award in New York Business for his medical devices, including his wearable technology. He has also been featured in several lay articles and also featured in videos and press releases. In 2018, he was a highlighted speaker in NYU’s TEDx “Re-Vision” Series. He is a member of American Medical Association, American College of Physicians, American Academy of PM\&R (AAPM\&R), Association of Academic Physiatry (AAP), and American Heart Association.
\end{IEEEbiography}
\EOD

\end{document}